\documentclass[twocolumn,tighten,times]{aastex63}

\hypersetup{colorlinks=true,linkcolor=blue,citecolor=blue,urlcolor=blue,}
\usepackage{verbatim} 
\usepackage{color,soul} 
\usepackage{graphicx} 
\usepackage[caption=false]{subfig}

\shorttitle{\sc Deep Learning VP Fitting}
\shortauthors{\sc Stemock \etal}

\input{iondefs.sty}

\begin{document}

\title{Deep Learning Voigt Profiles I. Single-Cloud Doublets}

\author[0000-0002-6434-4684]{Bryson Stemock}
\affiliation{Department of Astronomy, New Mexico State University, Las Cruces, NM 88003, USA}

\author[0000-0002-9125-8159]{Christopher W. Churchill}
\affiliation{Department of Astronomy, New Mexico State University, Las Cruces, NM 88003, USA}

\author{Avery Lee}
\affiliation{Department of Computer Science, New Mexico State University, Las Cruces, NM 88003, USA}

\author[0000-0002-1050-7572]{Sultan Hassan}
\altaffiliation{NASA Hubble Fellow}
\affiliation{Department of Astronomy, New Mexico State University, Las Cruces, NM 88003, USA}
\affiliation{Center for Cosmology and Particle Physics, Department of Physics, New York University, 726 Broadway, New York, NY 10003, USA}
\affiliation{Center for Computational Astrophysics, Flatiron Institute, 162 5th Ave, New York, NY 10010, USA}\affiliation{Department of Physics \& Astronomy, University of the Western Cape, Cape Town 7535,
South Africa}

\author[0000-0002-9009-6768]{Caitlin Doughty}
\affiliation{Department of Astronomy, New Mexico State University, Las Cruces, NM 88003, USA}
\affiliation{Leiden Observatory, Leiden University, PO Box 9513, 2300 RA, Leiden, The Netherlands}

\author{Rogelio Ochoa}
\affiliation{Department of Astronomy, New Mexico State University, Las Cruces, NM 88003, USA}

\begin{abstract}

Voigt profile (VP) decomposition of quasar absorption lines is key to studying intergalactic gas and the baryon cycle governing the formation and evolution of galaxies. The VP velocities, column densities, and Doppler $b$ parameters inform us of the kinematic, chemical, and ionization conditions of these astrophysical environments. A drawback of traditional VP fitting is that it can be human-time intensive.  With the coming next generation of large all-sky survey telescopes with multi-object high-resolution spectrographs, the time demands will significantly outstrip our resources.  Deep learning pipelines hold the promise to keep pace and deliver science digestible data products. We explore the application of deep learning convolutional neural networks (CNNs) for predicting VP fitted parameters directly from the normalized pixel flux values in quasar absorption line profiles. A CNN was applied to 56 single-component {\MgIIdblt} doublet absorption line systems observed with HIRES and UVES ($R=45,000$). The CNN predictions were statistically indistinct from a traditional VP fitter. The advantage is that once trained, the CNN processes systems $\sim\!10^5$ times faster than a human expert VP fitting profiles by hand. Our pilot study shows that CNNs hold promise to perform bulk analysis of quasar absorption line systems in the future. 
\end{abstract}



\section{Introduction}
\label{sec:intro}

Voigt profile (VP) fitting of quasar absorption lines has a long history as a vital tool for advancing our understanding of cosmic gaseous structures. VP fitting of {\Lya} forest lines \citep[e.g.,][]{morton72a, hu95, lu96, kirkman97, kim07, misawa07, danforth10, kim13, hiss18, garzilli20} has been key for constraints on the redshift clustering, column density distributions, and the temperatures and kinematics of the intergalactic medium (IGM). VP fitting has been crucial for constraining the D/H ratio in distant galaxies and thus cosmic baryon density \citep[e.g.,][]{burles98, tytler99}. Investigations into the cosmic evolution of fundamental physical constants, such as the fine structure constant rely heavily on VP fitting to quasar absorption lines spectra \citep[e.g.,][]{webb99, murphy17, bainbridge17}.

In {\Lya} absorption-selected systems, many galactic gas clouds in the circumgalactic medium (CGM) give rise to metal lines representing a range of ionization levels. VP fitting has served to provide the column density constraints for chemical-ionization models from which the densities, cloud structures, and metallicities can be measured \citep[e.g.,][]{bergeron86, bergeron94, peroux06, prochter10, lehner14, lehner16, lehner18}.  

Using {\CIV} absorption-selected systems, VP fitting has been used for constraining the kinematic, chemical, and ionization conditions and the cosmic evolution of IGM and CGM gas structures, including the nature of the ultraviolet ionizing background radiation \citep[e.g.,][]{morton72b, rauch96, songaila98, kim02, simcoe04, ryan-weber06, becker09, boksy15, cooper19, manuwal19}.  The high-ionization component of the CGM and IGM have also been extensively studied using the VP methodology as applied to {\OVI} absorption-selected systems \cite[e.g.,][]{simcoe02, simcoe04, danforth06, tripp08, muzahid12, johnson13, werk13, savage14, muzahid15, pointon19}, including those exhibiting {\NeVIII} absorption \citep[e.g.,][]{savage05}. And finally, the kinematic, chemical, and ionization conditions of low-ionization {\MgII} absorption-selected absorbers have also been studied using the powerful tool of VP fitting \citep[e.g.,][]{churchill97, rigby02, churchill03, lynch07, narayanan08, evans13, cooper19, churchill20}.  

Many VP fitting routines have been developed and applied to quasar absorption line systems \citep[e.g.,][]{vidal-madjar77, welty91, carswell91, fontana95, mar95, churchill97, carswell14, howarth15, bainbridge17, gaikwad17, liang17, krogager18, cooke19}. However, over the last few decades, the most commonly used fitting routine is {\sc VPfit} \citep{carswell14}.  

For complex systems, VP fitting is time intensive. For example, it required more than three years of human effort to fit $\simeq\! 420$ {\MgII} absorption-selected systems in $\simeq\! 250$ HIRES and UVES quasar spectra \citep{evans11, churchill20} and roughly eight years for \citep{boksy15} to fit $\simeq\! 200$ {\CIV} absorption-selected systems in nine HIRES spectra.  
One of the most human-intensive steps in the process is the creation of an initial guess VP model, in which the number of components, and the column densities, velocity centers, and Doppler $b$ parameters are estimated in a ``$\chi$-by-eye'' approach. The guess VP model is then used as a starting point for a least-squares fitting algorithm, which typically minimizes a $\chi^2$ statistic or maximizes a likelihood function. 

Deep learning artificial intelligence, such as an artificial neural network, holds potential to create breakthroughs in modern astronomy and cosmology via pattern recognition, clustering identification, scatter reduction, bias removal, anomaly detection, and the ability to efficiently simulate new data sets. These algorithms use training data sets to ``learn'' the defining characteristic properties within the data. Then, real-world data are presented to the network, which predicts the characteristic properties in the real data.  Successful applications include measuring galaxy star formation rates \citep[e.g.,][]{veneri19, simet21, euclid23, santos-olmsted23}, metallicities \citep[e.g.,][]{liew-cain21}, stellar masses, and redshifts \citep[e.g.][]{bonjean19, wu19, surana20}, galaxy cluster masses \citep[e.g.][]{ntampaka15}, cosmological parameters from weak lensing  \citep[e.g.][]{gupta18}, large-scale structure formation \citep[e.g.][]{he18}, identifying reionization sources \citep[e.g.][]{hassan19a} and the duration of reionization \citep[e.g.][]{la_plante19}, and constraining cosmological parameters \citep[e.g.][]{fluri19, ribli19, hassan20, matilla20, ntampaka20, ntampaka22, andrianomena23, bengaly23, lu23, novaes23, qiu23}. Recently, \citet{monadi23} applied Gaussian processes to SDSS DR12 quasar spectra to detect {\CIV} absorbers and measure their VP parameters. For a more complete description of the successful applications of machine learning to galaxy surveys and cosmology, we direct the reader to \citet{ntampaka19} and \citet{huertas-company23}.

Convolutional neural networks (CNNs) have also been successfully applied to quasar spectrum classification and quasar absorption line measurements.  \citet{pasquet-itam18} classified quasars in SDSS spectra and predicted their photometric redshifts with a 99\% success rate.  \citet{busca18} applied a deep learning CNN that identified a quasar sample 99.5\% pure and sub-classified broad absorption line (BAL) quasars with 98\% accuracy \citep[also see][]{guo19}. The 98\% success rates compare to human success rates; the difference is that a trained CNN accomplishes the task in several hours, whereas human-effort requires several years. \citet{parks18} trained a CNN that predicts the damped {\Lya} absorbers (DLAs) redshift and column density in un-normalized SDSS spectra with a reliability matching previous human-generated catalogs. \citet{cheng22} trained a CNN to predict the redshift, column density, and Doppler $b$ parameter of {\HI} absorbers in high resolution data. These works eradicated the human-intensive labor of continuum fitting, absorption line searching and identification, and VP fitting. However, to the best of our knowledge, nobody has employed a CNN to determine the VP parameters of metal absorbtion line systems in quasar spectra.

In this paper, we explore the deep neural network technique using a CNN to obtain VP models of absorption line spectra.  For this pilot study, we focus on the astrophysically common resonant fine-structure {\MgIIdblt} doublet.  We assume a single VP component and focus on the ability of the CNN to correctly predict the component velocity, column density, and Doppler $b$ parameter as constrained by absorption profiles of both members of the doublet.  We have the trained CNN make predictions for 56 real-world single-component {\MgII} absorbers observed with the HIRES \citep{vogt94} and UVES \citep{dekker00} spectrographs and we quantitatively compare the CNN predictions with previously obtained human VP fits to these {\MgII} absorbers \citep{churchill20}. 

In Section~\ref{sec:problem} we outline the challenges and describe our approach to the problem. In Section~\ref{sec:data}, we describe the real-world data set we use for bench marking the CNN.  The design and training of the CNN is described in Section~\ref{sec:methods} and the results are presented in Section~\ref{sec:results}. We discuss our findings in Section~\ref{sec:discussion} and summarize our concluding remarks in Section~\ref{sec:conclude}.


\section{Distilling The Problem}
\label{sec:problem}

Machine learning is the art of developing computer algorithms that learn to identify patterns in data by building flexible and generalized mathematical models of these data.  For ``supervised learning,'' the mathematical model is developed through an iterative training process in which inputs are mapped to outputs (called ``labels''); the supervised algorithms learn to build a unique function that can, with no further human intervention, be used to predict outputs associated with inputs the machine has never seen. 

There are various algorithms (e.g., artificial neural networks, decision trees, Bayesian networks, simulated and genetic annealing), each optimized for various types of problems.  Artificial neural networks are especially suited for problems in which the ability to generalize must be achieved from limited information; examples include predicting text from any human's unique hand writing or personal speech patterns, or translating languages from hand writing and/or speech.  These artificial neural networks are computational models inspired by the physiology of the brain. A convolutional neural network (CNN) is even more specifically suited for the analysis of data such as images, time-series data, or spectra, where the information in one measurement (pixel) is not independent of neighboring measurements (pixels).

Our goal is to explore how well a CNN can perform Voigt profile fitting on absorption line systems.  This is a classic, if challenging, regression problem that will likely be solved through steps of increasing complexity. Absorption line systems typically comprise multiple transitions from multiple ions, and the absorption profiles typically show a complex multi-component structure. Component blending and unresolved saturation effects can blanket information, and the severity of these issues depends on the resolution of the recording instrument \citep[e.g.,][]{savage91}. Furthermore, multi-phase ionization conditions give rise to absorption line systems in which absorption profiles exhibit velocity offsets between the low-ionization ions and the high ionization ions \citep[e.g.,][]{tripp08, savage14, sankar20}. The variations are countless. Moreover, the signal-to-noise ratio (S/N) of real-world spectra can vary dramatically as a function of spectral wavelength depending on the on-source exposure time, the total telescope throughput and wavelength dependent sensitivity of the spectrograph, and the spectral energy distribution of the source. This can yield a suite of absorption lines from a single system for which some of the absorption lines are recorded with a high S/N and others with a low S/N. 

For a CNN to navigate all of these nuances, it would need to be trained to recognize every possible permutation that manifests in real-world absorption line systems. This would suggest that the problem should be broken into smaller problems in order to make progress.  First, a CNN should be trained for only a single (or highly similar) telescope/instrument(s), as each yields data with unique characteristics, i.e., spectral resolution with a specific instrumental line spread function (ISF), data quantization and pixelization, and noise patterns. Second, our first explorations should be highly controlled. For example, transitions from a single ion should be tested to eliminate complexities due to multiphase ionization conditions.  Third, single-component absorption lines should be targeted to avoid complexity due to variations in line-of-sight gas kinematics that yield a great variety of multi-component profile morphology.  A highly focused exploration of this nature still needs to grapple with the spectrograph resolution and ISF, pixelization, varying S/N of the data, and the curve of growth behavior of absorption lines.  Successes under these simple conditions must be demonstrated before we embrace the greater complexities of absorption line spectra.

Our approach is to target single-component absorption lines from the {\MgIIdblt} fine-structure doublet as observed with the HIRES/Keck facility.  The {\MgII} ion is well studied in HIRES spectra, including extensive Voigt profile fitting of hundreds of systems \citep[e.g.,][]{churchill99, rigby02, churchill03, churchill20}.   Single-component {\MgII} absorbers are a scientifically interesting population of quasar absorption lines in their own right that have been extensively studied with high-resolution spectra  \citep[e.g.,][]{tytler87, petitjean90, churchill96, churchill97, churchill99, rigby02, prochter06, lynch07, narayanan07, narayanan08, evans11, matejek12, chen17, codoreanu17, mathes17, churchill20}

\section{Data}
\label{sec:data}

The {\MgII} absorption line systems were recorded in 249 HIRES \citep{vogt94} and UVES \citep{dekker00} quasar spectra obtained with the Keck and Very Large Telescope (VLT) observatories, respectively. The wavelength coverage of the spectra range from approximately 3,000--10,000~{\AA}.  The resolving power of both instruments is $R = \lambda/\Delta\lambda = 45,000$, or $\Delta v \sim 6.6$ {\kms}, and the spectra have $p=3$ pixels per resolution element. Further details on the data and data reduction and analysis of these the spectra can be found in \citet{evans11}, \citet{mathes17}, and \citet{churchill20}.  

The {\MgII} doublets were searched for and detected using the objective criteria of \citet{schneider93} as implemented in the code {\sc sysanal} \citep{churchill99}. All {\MgII}~$\lambda 2796$ detections must exceed a $5\sigma$ significance threshold while their {\MgII}~$\lambda 2803$ counterparts must exceed a $3\sigma$ significance threshold.  The spectral region over which an absorption profile is analyzed is determined by where the flux, $f_\lambda$,  in the wings on either side of the absorption profile becomes consistent with the continuum flux, $f^0_\lambda$.  This accounts for the local noise in the data by using the criteria that the per pixel equivalent width, i.e., $(1-f_\lambda/f^0_\lambda)\Delta \lambda$, where $\Delta \lambda$ is the pixel width, has become consistent with the standard deviation of these values in the surrounding continuum.  We call this spectral region the ``absorbing region.'' We bring this point to the reader's attention now as it will play an important role in the training of the CNN.

In these data, 422 {\MgII} absorption selected systems were fitted with Voigt profiles using the least-squares minimization code {\sc minfit}\footnote{\url{https://github.com/CGM-World/minfit}.} \citep{churchill97}, which iteratively eliminates all statistically insignificant components while adjusting the components until the least-squares fit is achieved. The latest version of the code and the ``fitting philosophy'' are described in \citet[][]{churchill20}. 

\begin{figure*}[htb!]
\figurenum{1} \centering
\fig{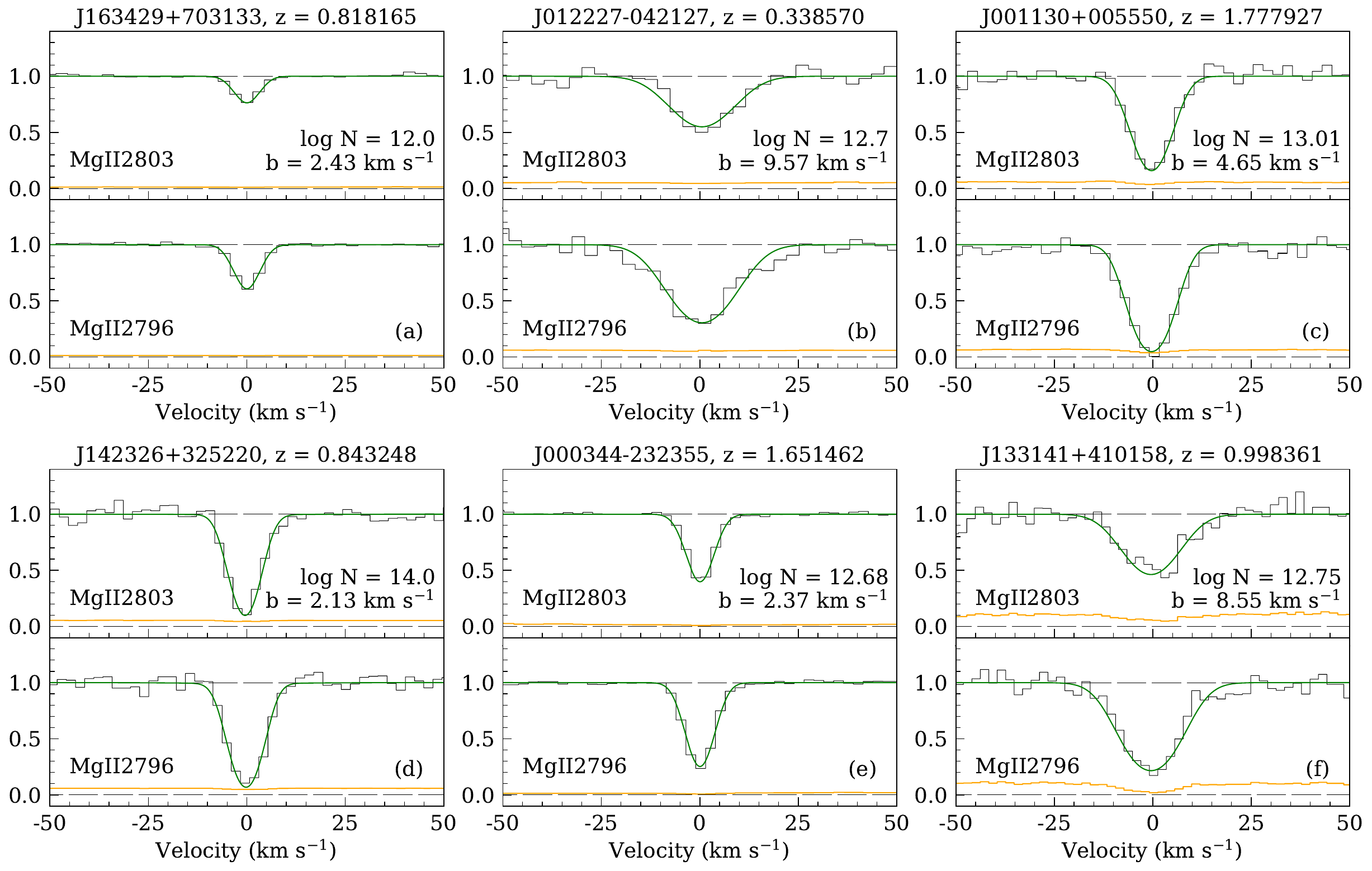} {0.9\textwidth}{} 
\vspace{-15pt}
\caption{
Characteristic systems from the set of 56 single-component {\MgII} systems observed by HIRES/Keck and UVES/VLT. The data are shown in black with green Voigt profile fits from \citet{churchill20}. Column densities and Doppler $b$ parameters from the Voigt profile fits are given in the {\MgII}~$\lambda 2803$ panel for each system. For the illustrated systems, the column densities range from $\log (N / {\rm cm}^{-2}) \simeq 12$ to $14$ and the $b$ parameters range from $b\simeq 2.1$ to $9.6$~{\kms}.
}
\label{fig:ex_obs}
\end{figure*}

Of the 422 systems, 56 were fitted with a single VP component. We hereafter refer to these single-component {\MgII} systems as ``single-cloud'' systems. For purpose of illustration, six representative single-cloud systems and their VP fits are shown in Figure~\ref{fig:ex_obs}.  These observed systems exemplify a range of absorption profile shapes and signal-to-noise ratios.  Each VP component is characterized by three fitted parameters, the component velocity center, $v$, the {\MgII} column density, $N$, and the Doppler $b$ parameter. The vertical dashed lines above the continuum indicate the absorbing regions over which these profiles are defined using the criterion described above.

\begin{figure*}[tb!]
\figurenum{2} \centering
\fig{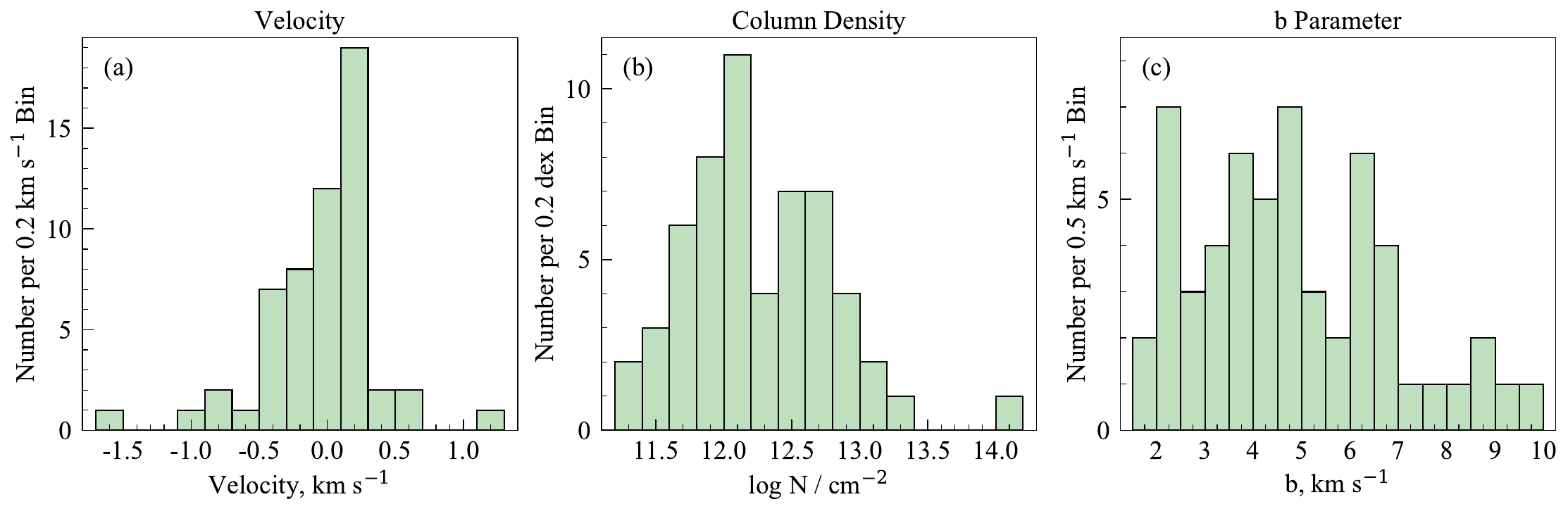}{0.9\textwidth}{}
\vspace{-15pt}
\caption{
Distributions of the VP parameters fit by \citet{churchill20} to the observed 56 single-component systems. Note that, since the absorption redshift (velocity zero point) is defined as the optical depth weighted median of the observed {\MgII}~$\lambda 2796$ profile using the flux values, the VP component velocity center can be non-zero. We concern ourselves with component velocities in this paper because this work serves as a springboard for analyzing more complex systems in which component velocity will be more important. Velocities range from $-1.51 \leq v \leq 1.11$~{\kms}, column densities range from $11.22 \leq \log (N/{\rm cm}^{-2}) \leq 14.12$, and $b$ parameters range from $1.62 \leq b \leq 9.57$~{\kms}.
}
\label{fig:obs_dists}
\end{figure*}

In Figure~\ref{fig:obs_dists}, we present the distributions of fitted VP parameters for the 56 single-cloud systems from \citet{churchill20}. In panel \ref{fig:obs_dists}(a), we show the VP component velocities, which range from $-1.51 \leq v \leq 1.11$~{\kms}. In panel \ref{fig:obs_dists}(b), we show the column densities, which range from $11.22 \leq \log (N/{\rm cm}^{-2}) \leq 14.12$. In panel \ref{fig:obs_dists}(c), we show the Doppler $b$ parameters, which range from $1.62 \leq b \leq 9.57$~{\kms}. Not shown in Figure~\ref{fig:obs_dists} is the signal-to-noise distribution of the continuum adjacent to the {\MgII}~$\lambda 2796$ profiles, which ranges from $9.5 \leq \textrm{S/N} \leq 90$ with one outlier at $\textrm{S/N} \sim 150$. We note that the VP component velocity is not precisely $v=0$~{\kms}. This is because the redshift of the {\MgII} profiles are defined as the optical depth median of the absorption profile \citep[see][]{churchill97}, which is computed directly from the flux decrements across the absorbing region of the {\MgII}~$\lambda 2796$ profile. This definition of the velocity zero-point is designed to establish the ``absorption redshift'' of a {\MgII} system even in the most kinematically diverse and complex absorption profiles.


\section{Methods}
\label{sec:methods}

To explore the ability of artificial intelligence to perform Voigt profile fitting, we employ a CNN for its aptitude for learning from data in which features are strongly correlated with neighboring features. This work focuses on two transitions, giving each training instance a shape of $2\times N_{\rm pix}$, where $N_{\rm pix}$ is the number of pixels in a single spectrum. Future work will include transitions from hydrogen, carbon, silicon, etc., and the CNN architecture provides the flexibility to grow in this way. Though CNNs are ignorant of atomic physics, ionization processes, and thermodynamics, they have a natural ability to ``see'' patterns within the data; as such, they learn the way the physics manifests in the data without learning the physics. Studies demonstrate very good ability of CNNs to outperform humans, even when the data are chaotic \citep{mandziuk02}.

The CNN is implemented using the {\sc Tensorflow} package v2.9.2 ~\citep{abadi15} and all CNN training and testing was completed using the New Mexico State University High Performance Computing cluster with 38 CPU and 16 GPU nodes, with total of 1,536 cores \citep{trecakov21}.

Our step-by-step iterative process of designing, training, and evaluating the CNN is as follows:

\begin{enumerate}
\item Design a CNN architecture for the deep learning of 2-dimensional arrays of flux in the first dimension and transition in the second dimension. 
\item Create a training set of synthetic {\MgII} absorption doublets in spectra with the characteristics of the HIRES/UVES instruments. These spectra must represent the resolution and pixelization, as well as the full range of S/N, column densities, $b$ parameters, and rest-frame velocity centers of the observed data. We normalize these parameters to a zero mean and unit standard deviation, i.e., a $N(1,0)$ normalization. Otherwise, the largest parameter will dominate the loss function calculation and limit the CNN performance.
\item Train the CNN to predict the VP parameters of single-cloud {\MgII} doublets using the aforementioned training spectra while withholding $\sim20\%$ of these doublets to be used to evaluate the CNN performance (known as the validation data).
\item Measure the accuracy and precision of the CNN predictions by having the CNN examine the validation data with known VP parameters. Given satisfactory results, test the CNN further using the 56 observed {\MgII} doublets.
\item After studying the results, redesign the CNN and/or training data as necessary to improve accuracy and precision. Employ a grid search to optimize the CNN hyperparameters. Repeat steps 1-5 as necessary.  
\item Study the sensitivity of the optimized CNN predictions to the adopted ISF, pixelization, and S/N.
\end{enumerate}

In the following subsections, we describe the CNN design and implementation and the construction of the training sample.  For a brief discussion of alternative algorithms and architectures, see Section~\ref{sec:discussion}. 

\begin{deluxetable}{ccc}[tbh]
\tablewidth{0pt}
\tablecaption{The Adopted CNN Architecture
\label{tab:cnn_architecture}}
\tablehead{
\colhead{Layer No.} & \colhead{Layer Function} &
\colhead{Output shape}
}
\startdata
1 & Input & ($226\times2$) \\[-4pt]
2 & 1D Convolutional Layer & ($226\times32$) \\[-4pt]
  & Batch Normalization & -- \\[-4pt]
  & ReLU Activation & -- \\[-4pt]
3 & 1D Convolutional Layer & ($226\times32$) \\[-4pt]
  & Batch Normalization & -- \\[-4pt]
  & ReLU Activation & -- \\[-4pt]
  & Max Pooling Layer & ($113\times32$) \\[-4pt]
  & Flattening Layer & ($3616\times1$) \\[-4pt]
4 & Fully Connected Layer & ($1000\times1$) \\[-4pt]
  & Batch Normalization & -- \\[-4pt]
  & ReLU Activation & -- \\[-4pt]
5 & Fully Connected Layer & ($800\times1$) \\[-4pt]
  & Batch Normalization & -- \\[-4pt]
  & ReLU Activation & -- \\[-4pt]
6 & 15\% Dropout Layer & -- \\[-4pt]
7 & Fully Connected Layer & ($100\times1$) \\[-4pt]
  & Batch Normalization & -- \\[-4pt]
  & ReLU Activation & -- \\[-4pt]
8 & Fully Connected Layer & ($3\times1$) \\
\enddata 
\end{deluxetable}

\subsection{CNN Design}
\label{subsec:design}

We present the adopted CNN architecture in Table~\ref{tab:cnn_architecture}. The {\MgII} absorption line data are prepared as a $2\times 226$ array, with the first dimension storing each of the two {\MgIIdblt} transitions and the second dimension storing the pixel flux values of the transitions.  The preparation of the data for the CNN is described in Section~\ref{subsec:training}. The data are then fed into two convolutional layers. Each convolutional layer is followed by
batch normalization and a Rectified Linear Units  \citep[ReLU, e.g.,][]{nair10} activation step. 

The convolutional layers use a kernel to sweep across the data, distilling information in local regions within the data. We adopted a $2\times1$ kernel for layer 2 and a $3\times1$ kernel for layer 3.  Convolutional layers employ a specified number of filters, and this gives the dimensionality (output shape) of the layer.  We adopted 32 filters in both convolutional layers.
Following the two convolutional layers, max pooling down samples the data, passing forward the most prominent, information rich features. After pooling, the data are collapsed, or flattened, into a 1D array. 

The remainder of the architecture comprises four fully connected layers.
Each fully connected layer is followed by batch normalization and a ReLU activation step. The fully connected layers start at 1,000 nodes (also referred to as the number of dense units) in layer 4, distill down to 800 in layer 5, and decrease again to 100 in layer 7.  Thus, the output shape of the data is progressively reduced.  Before the third fully connected layer, we employ a dropout layer (6). This combats overfitting by randomly deactivating 15\% of the nodes during the iterative learning process. The final layer (8) is a fully connected layer that consists of three nodes, corresponding the number of labels (VP component velocity, column density, and Doppler $b$ parameter).

Various other hyperparameters and functions also govern CNN performance. Importantly, the number of dense units (nodes) of each fully connected layer must be chosen.  The CNN processes the training set in subsets called batches. The batch size hyperparameter dictates how many instances (absorption doublets) the CNN processes before updating weights and biases. When the full training set (all instances) has been processed, this is called a training epoch.  We employed a stopping condition to determine the number of epochs the CNN will iterate. We adopted the stopping criteria that if three epochs pass without a decrease in loss of at least 0.05, the CNN will terminate training. In practice, the CNN trained for five epochs over the course of roughly 5 minutes.  The convergence of the CNN is controlled by a hyperparameter known as the learning rate. The learning rate scales the amount that the weights and biases are adjusted after each batch is processed. A loss function and optimizer must also be chosen. The loss function quantifies the difference between the CNN predictions and the true values (labels). The optimizer uses this information, as well as the learning rate, to modify the weights and biases after each batch is processed.  We adopted a learning rate of $10^{-4}$ and employed a mean squared error loss function coupled with the RMSprop optimizer\footnote{\url{https://keras.io/api/optimizers/rmsprop/}}.
For a thorough explanation of CNN hyperparameters and functionality, we refer the reader to \citet{erdmann21}.  

We conducted a controlled exploration to identify a combination of hyperparameters that yields superior performance.  Our explorations focused on three of the most important hyperparameters: the size of the convolution kernels, the number of filters in the convolutional layers, and the number of nodes (dense units) in the fully connected layers. To evaluate the performance of the CNN, we calculated $R^2$, the coefficient of determination, to compare the CNN predictions of the VP parameters against the true input VP parameters (the training labels). The value of $R^2$,
\begin{equation}
  R^2 = 1 - \frac{\sum (y_{predicted} - y_{target})^2}{\sum (y_{predicted} - \bar{y}_{predicted})^2} \, ,
\end{equation}
is the proportion of variance ``explained'' by the linear regression model, where $y_{predicted}$ is the predicted parameter from the CNN for some system, $y_{target}$ is the comparison value for that parameter for that system, and $\bar{y}_{predicted}$ is the mean of the predicted parameter for all systems. In the case of validation testing, $y_{target}$ is the true value used in the generation of the system. In the case of the observed data, $y_{target}$ is the parameter from the best-fit VP model as determined by {\sc minfit}.  We typically see $R^2 \in (0, 1)$, though highly skewed, flat predictions can result in $R^2 < 0$. A value of $R^2 \simeq 1$ indicates that the fraction of the variance unexplained in the model is vanishingly small, which constitutes a superior model.

We performed a grid search for hyperparameter selection to optimize CNN performance.  For each location on the search grid, the CNN was trained on $10^5$ systems and evaluated using both a validation data set with $10^4$ systems and 56 observed systems.  The construction of the training data is described in Section~\ref{subsec:training}.

\begin{figure*}[ht!]
\figurenum{3} \centering
\fig{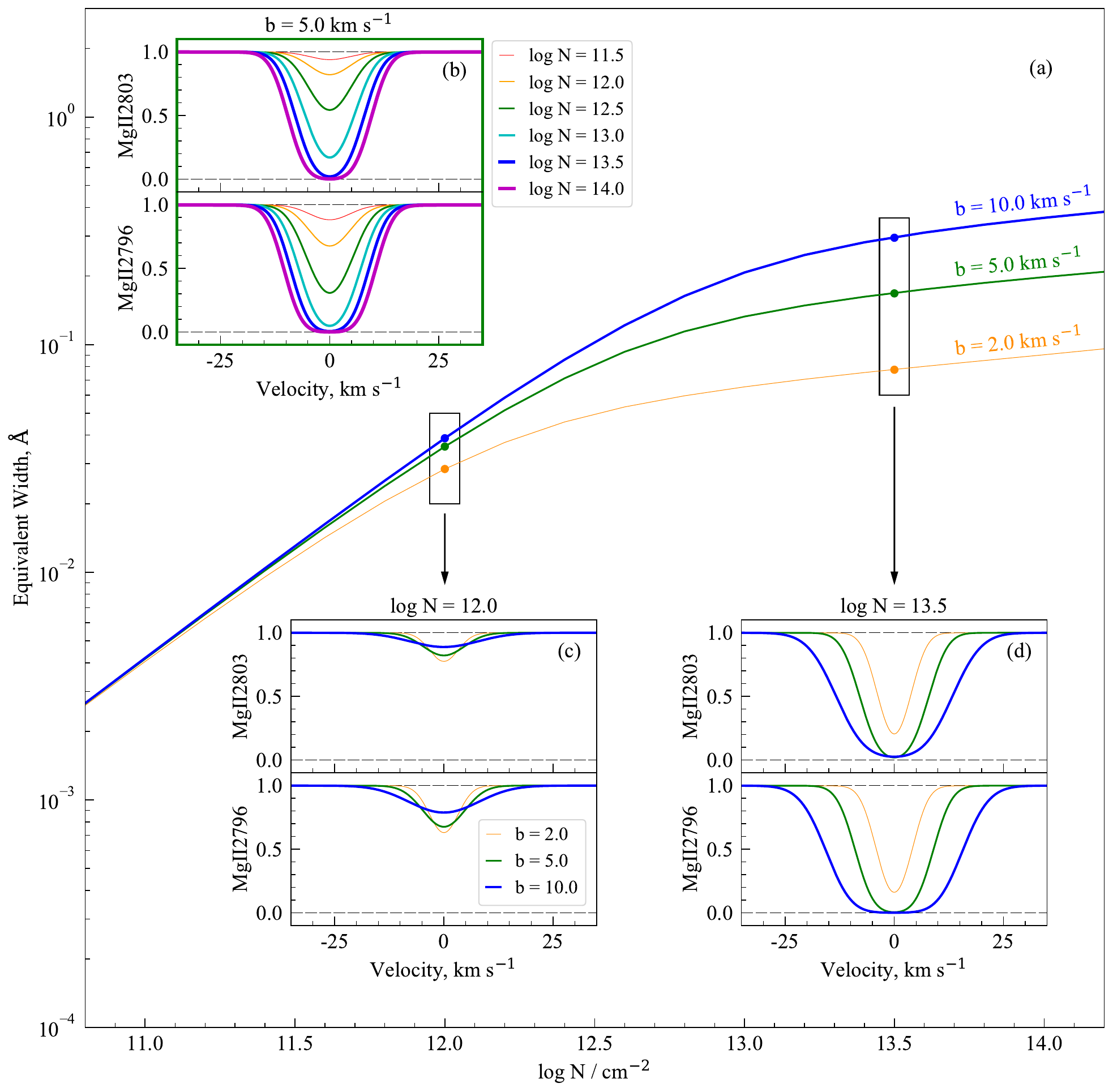}{0.7\textwidth}{}
\vspace{-15pt}
\caption{
A visual representation of the {\MgII} doublet profile shapes the CNN must learn and the underlying physics that connects them to the VP parameters. Panel (a) shows the curve of growth for three $b$ parameters: 2.0 km s$^{-1}$ (orange), 5.0 km s$^{-1}$ (green), and 10.0 km s$^{-1}$ (blue). Panel (b) demonstrates how the noiseless profiles of a $b = 5.0$~{\kms} system change given a range of column densities. Panels (c) and (d) show how the profiles of systems with column densities of $N=10^{12.0}$ and $10^{13.5}$~cm$^{-2}$, respectively, vary for each of the three $b$ parameters shown in Panel (a). The CNN must learn how $N$ and $b$ affect the shape of the data to accurately predict these parameters.
}
\vglue 0.15in
\label{fig:cog}
\end{figure*}

We first performed the CNN evaluations over a grid of hyperparameters for the convolutional layers (2,5) and then over a grid of hyperparameters for the fully connected layers (10,13,17). In each convolutional layer, we explored the size of the convolutional kernel over the grid $n\times 1$, where $n = 2, 3, 4, 5, 6$. For these layers, we also explored the number of filters, $m$, with $m = 2, 4, 8, 16, 32$. For each fully connected layer, we evaluated the CNN over a grid of the number of nodes (dense units) $M$, with $M \in  [100,1000]$ in steps of 100.  To assess which CNN architectures made superior predictions, we utilized the product of the $R^2$ values of $\log N$ and $b$, i.e., $R^2_N R^2_b$.  Those CNN with $R^2_N R^2_b \sim 1$ for which the learning rate and loss function indicated that the CNN was not overfitting were identified.  Though there is no ``best'' CNN architecture, we were able to identify a small subset of hyperparamters that yielded superior results. 
Further exploration would have involved grid searches of dropout rates (in layer 16), batch sizes, learning rates, etc. These efforts were deemed unnecessary given the near unity $R^2$ values we are able to achieve for the adopted CNN (see Section~\ref{sec:results}).

\subsection{The Training Data}
\label{subsec:training}

\begin{figure*}[ht]
\figurenum{4} \centering
\fig{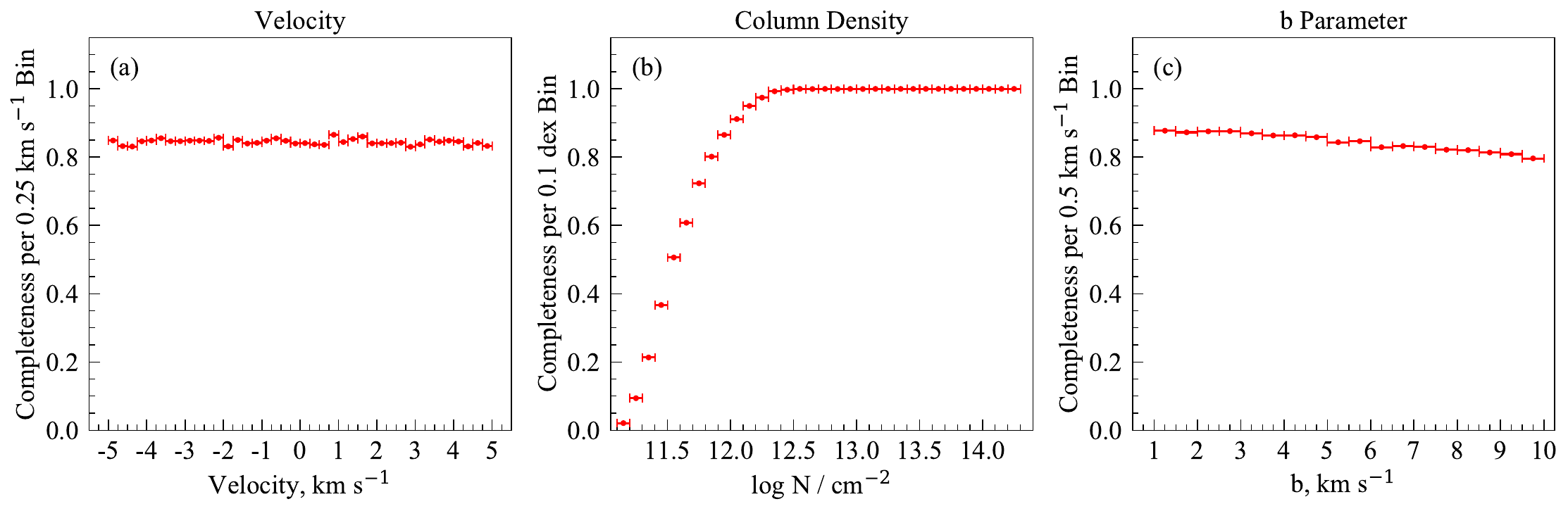}{0.95\textwidth}{}
\caption{
Completion curves of the simulated training set. Systems that were not detected at the $5\sigma$ and $3\sigma$ levels for {\MgII}~$\lambda 2796$ and {\MgII}~$\lambda 2803$, respectively, were removed from the training sample and deemed non-detections to accurately reflect the process of a human observer curating their database of systems. Note that the most frequently removed systems were those with low column densities or large Doppler $b$ parameters, which in concert would produce broad, shallow profiles more easily masked by random noise. Despite retaining only 360 of the original 6,250 systems with column densities less than $10^{11.3}$, the CNN was able to make accurate predictions for systems with the lowest column densities.
}
\vglue 0.15in
\label{fig:sim_dists}
\end{figure*}

The behavior of Voigt profiles are dictated by atomic and thermal physics as manifest in curve of growth.  We aim to teach the CNN these behaviors, which are illustrated in Figure~\ref{fig:cog}. Panel \ref{fig:cog}(a) illustrates the curve of growth for $b=2$, $5$, and $10$~{\kms}, a range typical of observed {\MgII} VP components \citep[][also see Figure~\ref{fig:obs_dists}]{churchill03,churchill20}. Panel \ref{fig:cog}(b) demonstrates how {\MgIIdblt} profile shapes vary with column density over the range $11.4 \leq \log (N/{\rm cm}^{-2}) \leq 14.0$ for $b=5$~{\kms}.  Similarly, panels \ref{fig:cog}(c,d) show how the profiles vary for $\log (N/{\rm cm}^{-2}) = 12.0$ and $\log (N/{\rm cm}^{-2}) = 13.5$ for  $b=2$, $5$, and $10$~{\kms}. Figure~\ref{fig:cog} also demonstrates the utility of teaching the CNN about both members of the {\MgIIdblt} doublet. For example, for $b=5$~{\kms} and $\log (N/{\rm cm}^{-2}) \geq 13.5$, the flux in the {\MgII}~$\lambda 2796$ line core becomes vanishingly small (saturates) and leverage on the column density dependence decreases dramatically. The broadening of the line wings carries  information, but the dependence of the profile shape and the line strength on column density is very weak (as can be seen by flattening of the green curve in Figure~\ref{fig:cog}(a)).  When noise and pixel discretization are accounted, it is not trivial to decouple broadening caused by additional column density from that caused by an increase in $b$. The leverage is enhanced by inclusion of the {\MgII}~$\lambda 2803$ line, which does not saturate in a different manner for the same column density and $b$ parameter. 

Each absorption system in the training set is defined by four quantities: the S/N, the velocity, the column density, and the Doppler $b$ parameter. Based on the ranges of these parameters in the observed data, given in Section~\ref{sec:data}, we adopt prior range for our training set of $9 \leq \hbox{S/N} \leq 90$, $-5.0 \leq v \leq 5.0$ {\kms}, $11.1 \leq \log (N/{\rm cm}^{-2}) \leq 14.3$, and $1.0 \leq b \leq 10.0$ {\kms}. Our velocity prior range is significantly wider than the observed data range of $-1.51 \leq v \leq 1.11$ {\kms} to avoid velocity variations that are purely on the sub-pixel level ($\Delta v_{\rm pix} < 2.22$~{\kms}).

We perform Latin Hypercube Sampling using the {\tt PyDOE} python package\footnote{\url{https://github.com/tisimst/pyDOE}.} to generate $10^5$ ordered quadruplets of (S/N, $v$, $\log N$, $b$). This method provides a uniform density of points with added local randomness, allowing us to generate multiple unique data sets that sample the same parameter space. Thus, we can ensure the CNN properly learns to generalize trends in Voigt profiles as opposed to learning a particular data set really well, a problem commonly referred to as ``overfitting.''

For these $10^5$ ordered quadruplets, each goes through the following process:

\begin{enumerate}
\item The flux values for the {\MgIIdblt} transitions are generated (described further below). These profiles are  convolved with the ISF, pixelized, and noise is added.  
\item The absorption lines are run through the detection software and the absorbing regions for each transition are identified. If the {\MgII}~$\lambda 2796$ and {\MgII}~$\lambda 2803$ lines are not detected at the $5\sigma$ and $3\sigma$ levels, respectively, the system is removed from the sample. This exactly emulates the manner in which real-world data are included in the observational sample.
\item {\MgIIdblt} flux values are recorded and concatenated into a $226 \times 2$ array for CNN training.
\end{enumerate}

In Step~1, the doublets are generated in units of relative flux using Voigt profiles, each defined by its velocity center, column density, and $b$ parameter. Details of how the spectra are generated are explained in \citet{churchill15}. As we are working with a single ion, one cannot decouple the thermal and non-thermal components to the $b$ parameter, so we instead use the total $b$ parameter. The Voigt profiles are convolved with a Gaussian ISF with a full-width half-maximum resolution element defined as $\Delta \lambda = \lambda/R$. To accurately emulate HIRES/UVES spectra, we adopt $R=45,000$. This was the resolution adopted for the VP fitting software that fitted the real-world data. In Appendix~\ref{app:resolution}, we explore the sensitivity of the CNN to the adopted resolution.  The convolved spectra are then pixelated as defined by the factor $p$, the number of pixels per resolution element, according to $\Delta \lambda = p \Delta \lambda_{\rm pix}$, where $\Delta \lambda_{\rm pix}$ is the wavelength extent of a pixel.  We adopt $p=3$, the values appropriate for the HIRES and UVES spectrographs. Through the relations $R= \lambda/\Delta \lambda = \lambda/(p\Delta \lambda_{\rm pix}) = c/(p\Delta v_{\rm pix}$), we see that the velocity width of an individual pixel is $\Delta v_{\rm pix}=2.22$~{\kms}. Finally, we add Gaussian noise in each pixel by generating random deviates weighted by the Gaussian probability distribution function with $\sigma = (\textrm{S/N})^{-1}$ in the continuum. In the absorption line we account for the read-noise and reduced Poisson noise \citep[see][]{churchill15}, again, to ensure that we emulate the characteristics of real-world data. 

\begin{figure*}[htb!]
\figurenum{5} \centering
\fig{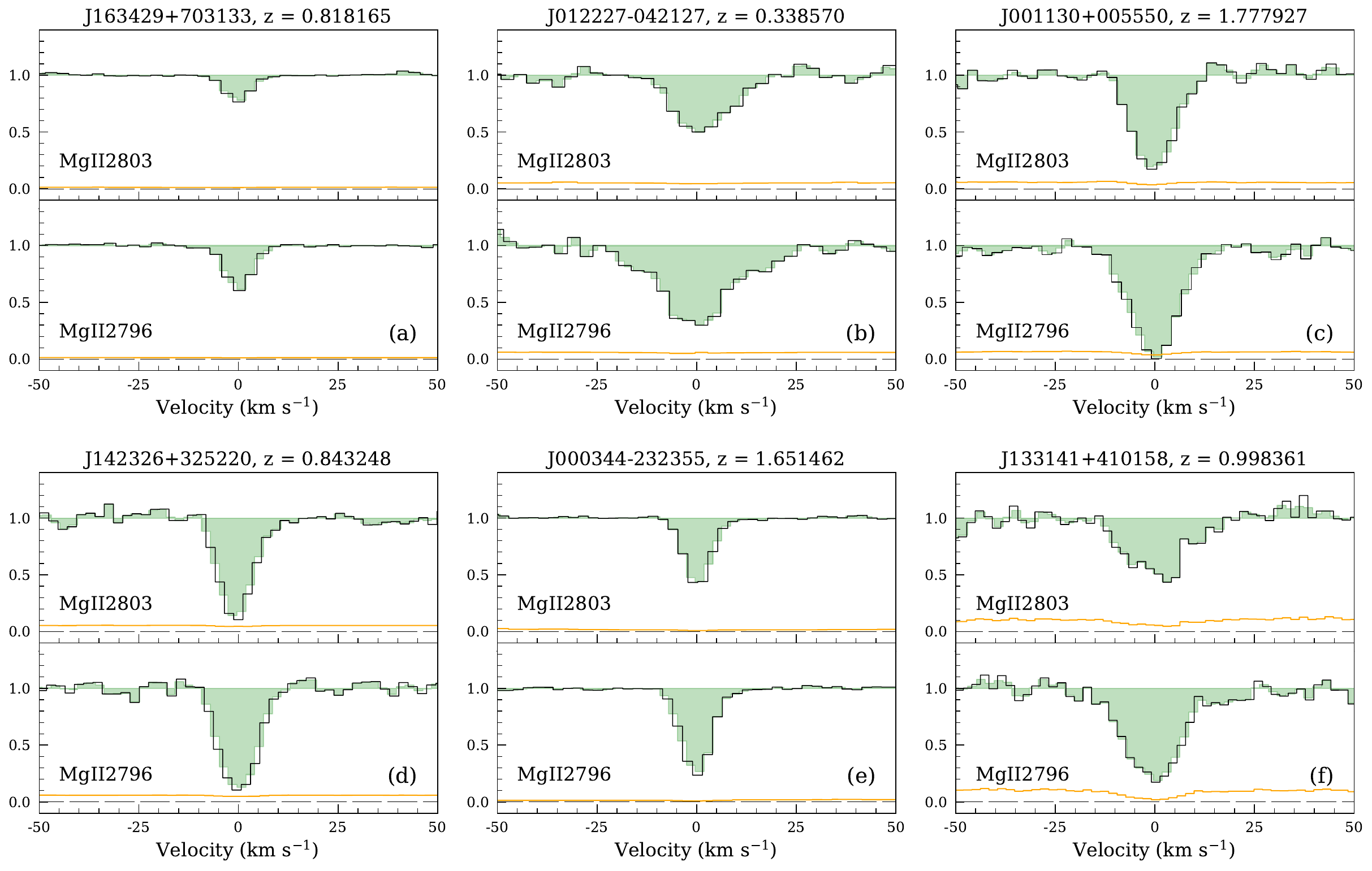}{0.95\textwidth}{}
\vspace{-15pt}
\caption{
Examples of preparation of the real-world data using the same {\MgII} doublets illustrated in Figure~\ref{fig:ex_obs}. The preparation of the data consists of rebinning the pixels to match those of the training data. Black histograms are the flux values prior rebinning, and the green shaded regions are the prepared data, using flux conservation. Rebinning is necessary because the CNN is given only flux versus pixel position, meaning that the velocity in each pixel must identically map with pixel position for every system.
}
\vglue 0.15in
\label{fig:rebinning}
\end{figure*}

In Step 2, roughly $16$\% of the original $10^5$ systems were discarded as non-detections. Figure~\ref{fig:sim_dists} gives the fraction of systems in each bin in the training set divided by the number of systems in each bin in the original $10^5$ sample. In practice, low column density, broad (high $b$ parameter) profiles are ``shallow'' and are more easily lost in the noise.  This trend is apparent in Figure~\ref{fig:sim_dists}(b,c), which shows that detectability drops dramatically for $\log (N/{\rm cm}^{-2}) < 12$ with a higher percentage undetected for higher $b$ parameters. However, there are still a sufficient number of systems with these lower column density and higher $b$ parameters for the CNN to learn these profiles without biasing results.

In Step~3, the spectrum of each transition is stored in an individual 1-dimensional spectral segment of 226 pixels, which occupy a velocity range of approximately $\pm 250~${\kms}. Every transition uses this fixed velocity grid to ensure the CNN understands velocity positioning despite seeing only a vector of flux values. We experimented with smaller velocity windows of $\pm 100~${\kms} and $\pm 50~${\kms} to reduce the number of extraneous pixels given to the CNN, but found that this did not improve the CNN performance.

\subsection{Preparing Real World Data for the CNN}
\label{subsec:obs_data_prep}

For each absorption system, the CNN sees $2\times226$ array of flux values.  The first dimension comprises the two transitions and the second comprises the 226 pixels.  At no point is the CNN directly informed about the velocity corresponding to a given pixel; the CNN learns the mapping from pixel to velocity during the supervised learning. Therefore, when applying the CNN to real-world data it is necessary that the mapping from pixel to velocity in the real-world data is identical to the mapping in the training data. We thus, bin the real-world data to have a constant pixel velocity width equal to that of the training data while enforcing the first pixel has the same velocity as the first pixel of the training data.  For deep learning, it is common to prepare the data.  We call our preparation ``rebinning.''

For rebinning, we invoke flux conservation.  In Figure~\ref{fig:rebinning}, we show examples of prepared data.  The systems shown here are the same as shown in Figure~\ref{fig:ex_obs}.  The green shaded regions are the prepared data after rebinning and the black histograms are the original observed data. Residuals between the prepared data and the original data are typically smaller than the original uncertainties in the flux values. 


\section{Results}
\label{sec:results}

After teaching the CNN, we evaluated it in two ways.  For the first, we compare the CNN performance to {\sc minfit} using ``withheld'' training data, which comprise $\sim\! 10^4$ simulated doublets not included in the training of the CNN. These data are also known as the validation set.  For the second, we assess the CNN ability to predict VP parameters for the prepared real-world data by comparing to the {\sc minfit} results on these data.

\subsection{Method Comparison Using Simulated Data}
\label{subsec:sim_data}

\begin{figure*}[b!t]
\figurenum{6} \centering
\fig{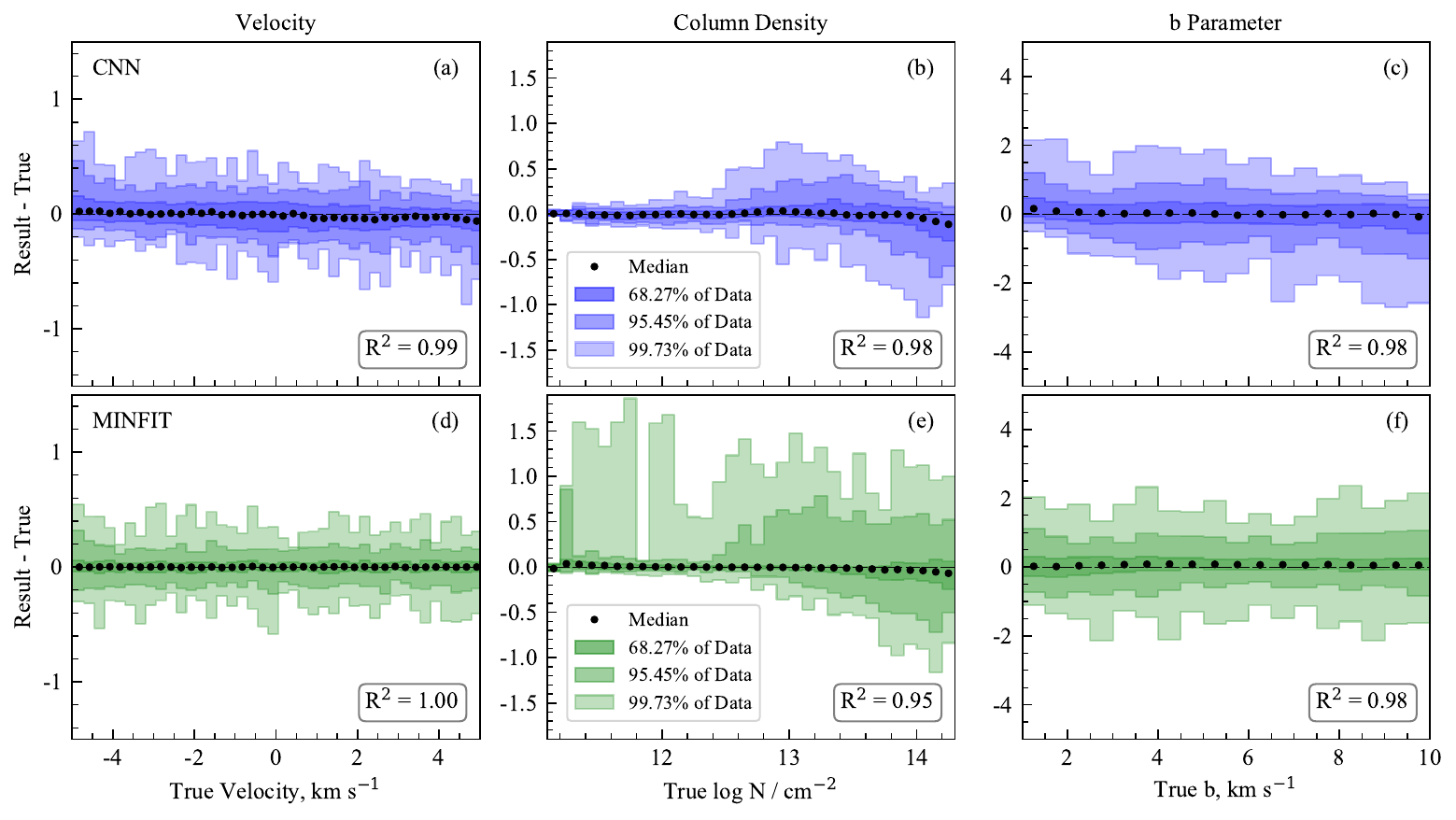}{0.9\textwidth}{}
\vspace{-15pt}
\caption{
Residuals of the velocities, column densities, and $b$ parameters for (a,b,c) the predictions of the CNN, and (d,e,f) the results of the least-square VP fitter for the validate data set (roughly $10^4$ systems).  We display the results in bins of 0.25 km s$^{-1}$ for velocity, 0.1~dex for column density, and 0.5~{\kms} for the $b$ parameter. Velocity residuals are plotted in units of pixels, where 1~pixel represents 2.2~{\kms}. Black dots indicate the median residual in a given bin and the lightest, medium, and darkest shaded regions contain 99.73\%, 95.45\%, and 68.27\% of the residuals, respectively. Both methods yield comparable results.}
\vglue 0.2in
\label{fig:method_residuals}
\end{figure*}

We compare two methods to recover the input velocities, column densities, and Doppler $b$ parameters used to generate the validation set. The first method, dubbed ``CNN,'' is the trained CNN predictions. The second method, referred to as ``{\sc minfit},'' employs the VP fitting code used in \citet{churchill20} (described in Section~\ref{sec:data}). We provide the true VP parameters as an initial model for {\sc minfit} least-squares fitting. This represents a best-case scenario for the traditional process of a human manually generating initial models and then applying {\sc minfit}.

Both the CNN predictions and the {\sc minfit} results are displayed in Figure~\ref{fig:method_residuals}.  Black dots represent the median residual as a function of the ``true value.''  For the CNN predictions (top row, blue data), the plotted values are the residuals between the CNN predictions and the true values used to generate the withheld data.  For the {\sc minfit} results (bottom row, green data), the plotted values are the residuals between the VP fits and the true values used to generate the withheld data.  The shading of the residuals provides the distribution of the residuals in terms of 68\% ($1\sigma$), 95\% ($2\sigma$), and 99\% ($3\sigma$) area contained within the distribution.  Note that the velocity residuals are sub-pixel in size, so we have plotted the residuals in units of pixels to simplify interpretation. 

\begin{figure*}[tb]
\figurenum{7} \centering
\fig{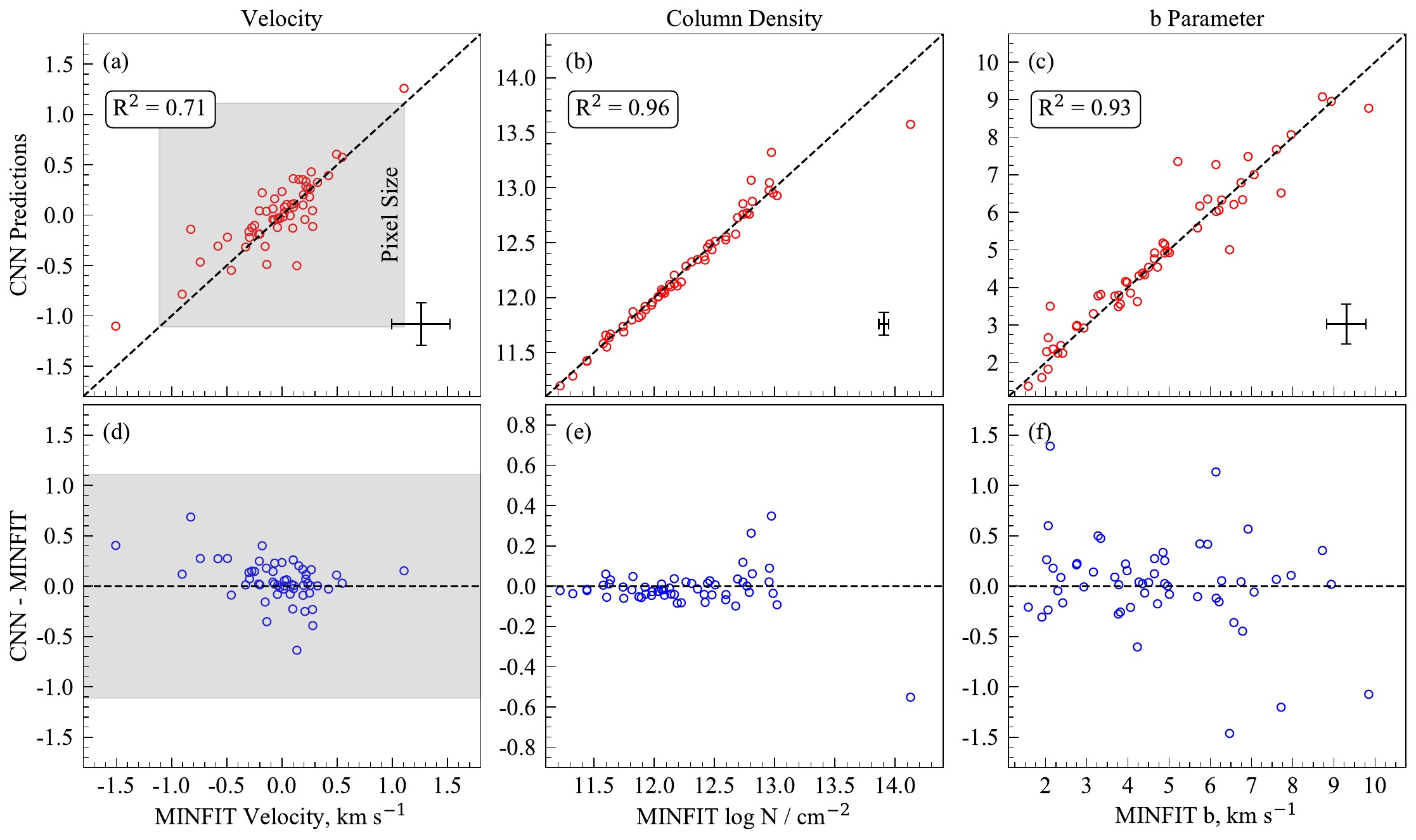}{0.9\textwidth}{}
\vspace{-15pt}
\caption{(a,b,c) The CNN predictions for VP component (a) velocity, (b) column density, and (c) Doppler $b$ parameter for the 56 observed single-cloud {\MgII} absorbers from \citet{churchill20}.  In the lower right corner of each panel, the RMS errors in the CNN predictions (vertical bar) and in the fitted parameters (horizontal bar) are given.  (d,e,f) The scatter plots of the CNN errors as a function of the {\sc minfit} parameter values. In panels (a) and (d) the width of a single pixel is shown as a shaded region. Note that the velocity zero point of the absorption lines is based on the median optical depth \citep[][]{churchill97} so that {\sc minfit} VP velocities of single-component absorbers can be non-zero. This figure shows general agreement between both methods on real-world data.}
\vglue 0.15in
\label{fig:cnn_obs}
\end{figure*}

In Figure~\ref{fig:method_residuals}(a,d), we find that the CNN and {\sc minfit} have $R^2>0.99$ when recovering rest-frame velocities. For both the CNN and {\sc minfit}, the residuals are a small fraction of a pixel. Furthermore, the distribution in the residuals are consistent with being flat (showing no skew) as a function of velocity. Roughly 99\% of the CNN predicted and VP fitted values reside within $\pm 0.2$ of a pixel, corresponding to $\pm 0.44$~{\kms}. 

In Figure~\ref{fig:method_residuals}(b,e), we see that column densities are predicted with $R^2=0.98$ for the CNN, whereas {\sc minfit} yields $R^2=0.95$.  Here the CNN has shown superiority to the traditional VP fitter in that the spread in the distribution of the residuals is substantially narrower around the mean values.  For $\log (N/{\rm cm}^{-2}) \sim 13.5$ the residuals from both the CNN and {\sc minfit} tend to slightly skew towards larger $N$, whereas for the CNN they slightly skew towards smaller $N$ for $\log (N/{\rm cm}^{-2}) \sim 14$. However, even in this regime of high column density, the residuals from the CNN are similar in magnitude to those from {\sc minfit}.

In Figure~\ref{fig:method_residuals}(c,f), we see that the Doppler $b$ parameters for both the CNN and {\sc minfit} both have have $R^2 = 0.98$.  The spreads in the distributions of residuals are highly similar.  However, there is a slight skew in the distribution of residuals for the CNN such that for narrower lines the residuals skew to larger $b$ and for broader lines the residuals skew to smaller $b$. Overall, the CNN predicted $b$ parameters are comparable to those of the traditional least-square VP fitter using the validation data set.

\subsection{Application to Real Data}
\label{subsec:real_data}

The ultimate goal is to have the CNN accurately predict the VP parameters of real-world data. It is important to remember that we do not know the the values of the ``true'' VP parameters for the observed data\footnote{This statement is made based on the assumption that real-world absorption lines arise from gas environments that manifest a true Voigt profile. By utilizing single component absorption lines that generally appear symmetric for our study, we feel we can comfortably embrace this assumption.}; we can only compare the predictions of the CNN to the parameters obtained from traditional methods, i.e., VP fitters.  The observed data were originally fit using {\sc minfit} by \citet{churchill20}.  However, some of these {\MgII} systems had accompanying absorption lines from {\FeII} transitions and/or the {\MgI}~$\lambda 2852$ transition.  Since we trained the CNN for the {\MgII} doublet lines only we refit these systems using only the {\MgIIdblt} transitions.  This allows us to eliminate possible systematic effects due to the influence of the kinematic structure of other absorption profiles and/or possible statistical effects due to the influence of the noise characteristics of the other transitions on the least squares fitting function.  For these ``refits,'' we adopted the human generated initial-guess VP models of  \citet{churchill20} for the {\MgII} profiles.

In Figure~\ref{fig:cnn_obs}, we present the CNN predictions versus the {\sc minfit} results for the observed data.  The direct comparison of the VP parameters is shown in panels \ref{fig:cnn_obs}(a,b,c). If we adopt the {\sc minfit} values as the ``true'' or benchmark values, then the ``errors'' of the CNN predictions can be computed from  $\Delta a = a_{\hbox{\tiny CNN}} - a_{\rm fit}$, where $a$ represent one of the VP parameters, $v$, $\log N$, or $b$. The RMS errors of the VP parameters are shown as the error bars in the lower right regions of panels \ref{fig:cnn_obs}(a,b,c), where the vertical error bar represents the CNN prediction errors and the horizontal bar represents the {\sc minfit} fitting errors from the covariance matrix of the least squares fitter.  A scatter plot of the CNN errors is shown in panels \ref{fig:cnn_obs}(d,e,f). The $R^2$ values for the regression model between the CNN predictions and the {\sc minfit} results are $R^2=0.71$ for velocities, $R^2=0.96$ for column densities, and $R^2=0.93$ for $b$ parameters.  

\begin{figure*}[htb]
\figurenum{8} \centering
\fig{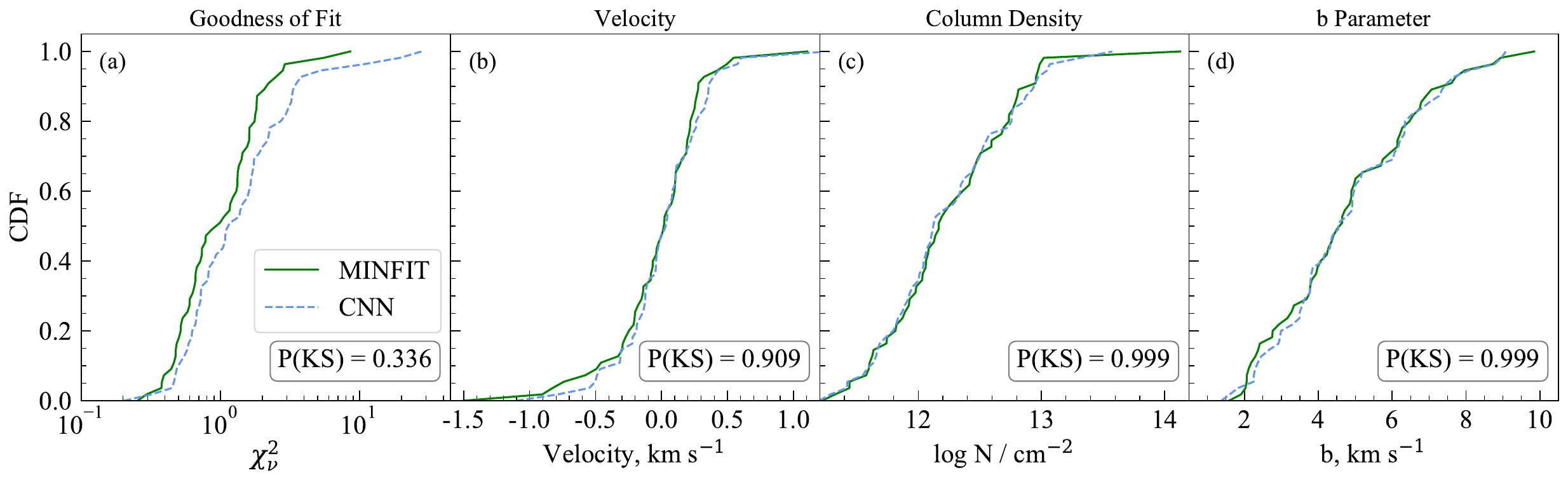}{0.9\textwidth}{}
\vspace{-15pt}
\caption{
Cumulative distribution functions for the 56 observed systems comparing the CNN predictions (blue, dashed) and {\sc minfit} results (green, solid). (a) The $\chi^2_\nu$ values for the VP profile models. (b) The VP component velocities. (c) The VP component column densities. (d) The VP $b$ parameters.  The KS probabilities, $P({\rm KS})$, that the two distributions are drawn from the same distribution are given, showing that in all cases, we cannot reject the null hypothesis that the distributions of results are indistinguishable. 
}
\vglue 0.15in
\label{fig:cdfs}
\end{figure*}

The $R^2$ value of 0.71 for the CNN velocity predictions might suggest that the CNN is not as effective in predicting VP velocities for the real data as it is at predicting column densities and $b$ parameters. However, the RMS error in these predictions, 0.210~{\kms}, is less than 10\% of a pixel velocity width, 2.22~{\kms}. The RMS error in the CNN predictions is less than the RMS error of the {\sc minfit} fitted velocities.  This suggests that, for the resolution, pixelation, and S/N range of the observed HIRES and UVES data, the precision of VP velocities is no better than 10\% of a pixel regardless of the method of VP parameter estimation.

The $R^2$ values of 0.96 and 0.93 for the column densities and Doppler $b$ parameters, respectively, indicate small residuals about the regression model. In fact, the RMS errors in the CNN predictions for these parameters are nearly equal to the errors in the {\sc minfit} parameters.  Interestingly, patterns in the errors shown in Figure~\ref{fig:cnn_obs}(e,f) are suggestive of the patterns seen in 
Figure~\ref{fig:method_residuals}(b,c).  For $\log (N/{\rm cm}^{-2}) \sim 13$ the CNN errors slightly skew towards larger $N$. And for the single system at $\log (N/{\rm cm}^{-2}) \sim 14$, the CNN error is negative, consistent with the skew toward smaller $N$ in this regime of column density.  A similar trend is seen for the $b$ parameters in that the CNN predictions for the broader lines can skew toward narrower lines.

The {\MgIIdblt} absorption profiles of the sole high-column system in the observed sample, which has $\log (N_{\rm fit}/{\rm cm}^{-2}) \!=\! 14.12\pm 0.70 $ and $b_{\rm fit} \!=\! 2.06\pm0.42$~{\kms}, are unresolved and saturated.  The signal-to-noise of the data is $\hbox{S/N} \!\sim\! 20$. The profiles are very narrow and deep and firmly reside on the flat part of the curve of growth (see Figure~\ref{fig:cog}).  The CNN predictions is $\log (N_{\hbox{\tiny CNN}}/{\rm cm}^{-2}) = 13.58$ and $b_{\hbox{\tiny CNN}} = 2.66$~{\kms}.  Statistically, the CNN predictions for the column density falls within the $1\sigma$ of the {\sc minfit} measurement and the CNN predictions for the $b$ parameter falls within $1.4\sigma$ of the {\sc minfit} measurements.  Visual inspection of the over plotted absorption profile synthesized from the CNN and {\sc minfit} VP parameters shows that they both accurately model the absorption.  The reduced chi-square statistic computed over the absorbing pixels for this system using the spectral models generated by each method shows that the CNN model yields $\chi^2_\nu = 1.39$ while the {\sc minfit} model yields $\chi^2_\nu = 0.74$ for the doublet.  Unfortunately, we have only one system in the real-world data sample that is highly unresolved and resides firmly on the flat part of the curve of growth for testing the CNN.  However, we note that the exercises conducted to generate Figures~\ref{fig:method_residuals}(b,c) and \ref{fig:method_residuals}(e,f) demonstrated that both the CNN and the traditional least-squared VP fitter struggle in this regime.

\begin{figure*}[htb!]
\figurenum{9} \centering
\fig{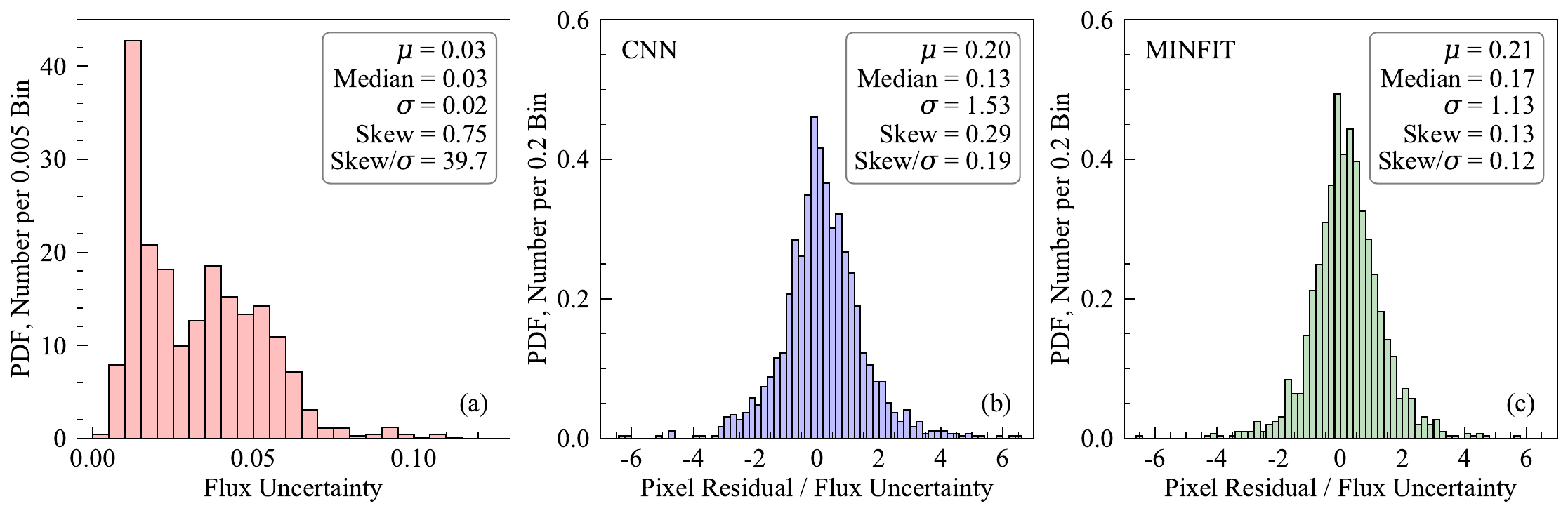}{0.9\textwidth}{}
\vspace{-15pt}
\caption{Panel (a): The distribution of uncertainties in the absorbing pixels.
The signal-to-noise, S/N, in a pixel is the inverse of its relative flux uncertainty, thus, the median signal-to-noise in the observed absorption profiles is $\langle \textrm{S/N} \rangle = \mu^{-1} = 33$. Panel (b): The distribution of the ``goodness of fit'' of the Voigt profile predicted by the CNN. The goodness of fit is defined as the residual between the Voigt profile and the observed absorption profile in each pixel divided by its relative flux uncertainty. Panel (c): The distribution of the ``goodness of fit'' of the Voigt profile fitted by {\sc minfit}. All values in this figure are quoted in terms of continuum-normalized flux. The CNN models produce more accurate results with slightly higher dispersion and skew.
}
\label{fig:flux_residuals}
\end{figure*}

To further assess the CNN, in Figure~\ref{fig:cdfs}(a) we present the $\chi^2_\nu$ statistics for the absorption line models generated from the VP parameters. 
In Figure~\ref{fig:cdfs}(b,c,d), we present the cumulative distribution functions (CDFs) of the VP velocities, column densities, and $b$ parameters.  We conducted two-sample Kolmogorov–Smirnov (KS) tests to differentiate whether the CNN predictions and the {\sc minfit} results are consistent with being drawn from the same underlying distributions.  To rule out that the CNN and {\sc minfit} distributions are drawn from the same distribution at the 99.97\% confidence level ($3\sigma$) or higher, we would require a KS probability $P({\rm KS}) \leq 0.0027$.  For the column densities and $b$ parameters, we obtain $P({\rm KS}) \simeq 1$, indicating a high level of confidence that the two distributions represent the same underlying distribution.  For the velocities, we obtain $P({\rm KS}) \simeq 0.9$, also indicating the two distributions represent the same underlying distribution.  For the ``goodness of fit,'' as quantified by the $\chi^2_\nu$ statistic, we obtain $P({\rm KS}) = 0.34$.  Whereas the $\chi^2_\nu$ appears to be somewhat normally distributed around unity for the {\sc minfit} results, it appears somewhat skewed toward larger values for the CNN predictions. This would indicate that the variance in the fit of the VP profile model is slightly systematically larger than the variance uncertainties in the flux values. Still, the KS statistic does not indicate a significant difference in the goodness of the fit between the CNN predictions and the {\sc minfit} results.

In Figure~\ref{fig:flux_residuals}(a), we present the distribution of uncertainties in the absorbing pixels quoted as the relative flux uncertainty, $\sigma_\lambda/f^0_\lambda$, where $f^0_\lambda$ is the continuum flux.  The statistical descriptors of the distribution is provided in the legend.  In Figure~\ref{fig:flux_residuals}(b,c), we present the distribution of the ratio of the pixel residual between the VP model and the flux value to the flux uncertainty,
\begin{equation}
  \frac{\hbox{pixel residual}}{\hbox{flux uncertainty}} =   \frac{Y_\lambda - f_\lambda/f^0_\lambda}{\sigma_\lambda / f^0_\lambda} \, ,
\label{eq:normalized-flux-residuals}
\end{equation}
where $Y_\lambda$ is the VP model evaluated at the pixel with wavelength $\lambda$. We call this quantity the uncertainty normalized model residuals.
The statistical descriptors of the distribution of normalized model residuals for the CNN predictions and the {\sc minfit} results are highly consistent with one another.  Both have a mean of $\mu \simeq 0.2$ and median of $\simeq 0.15$.  Though the dispersion of the CNN predictions is slightly broader than that of the {\sc minfit} results, both are quite narrow and highly symmetric with $k/\sigma \simeq 0.1$--0.2, where $k$ is the skew. A KS test comparing the distributions presented in panels \ref{fig:flux_residuals}(b) and \ref{fig:flux_residuals}(c) yields $P({\rm KS})= 0.117$, indicating that, based on the normalized model residuals, we cannot reject the null hypothesis that these results represent the same population of VP models.

\section{Discussion}
\label{sec:discussion}

\begin{figure*}[htb!]
\figurenum{10} \centering
\fig{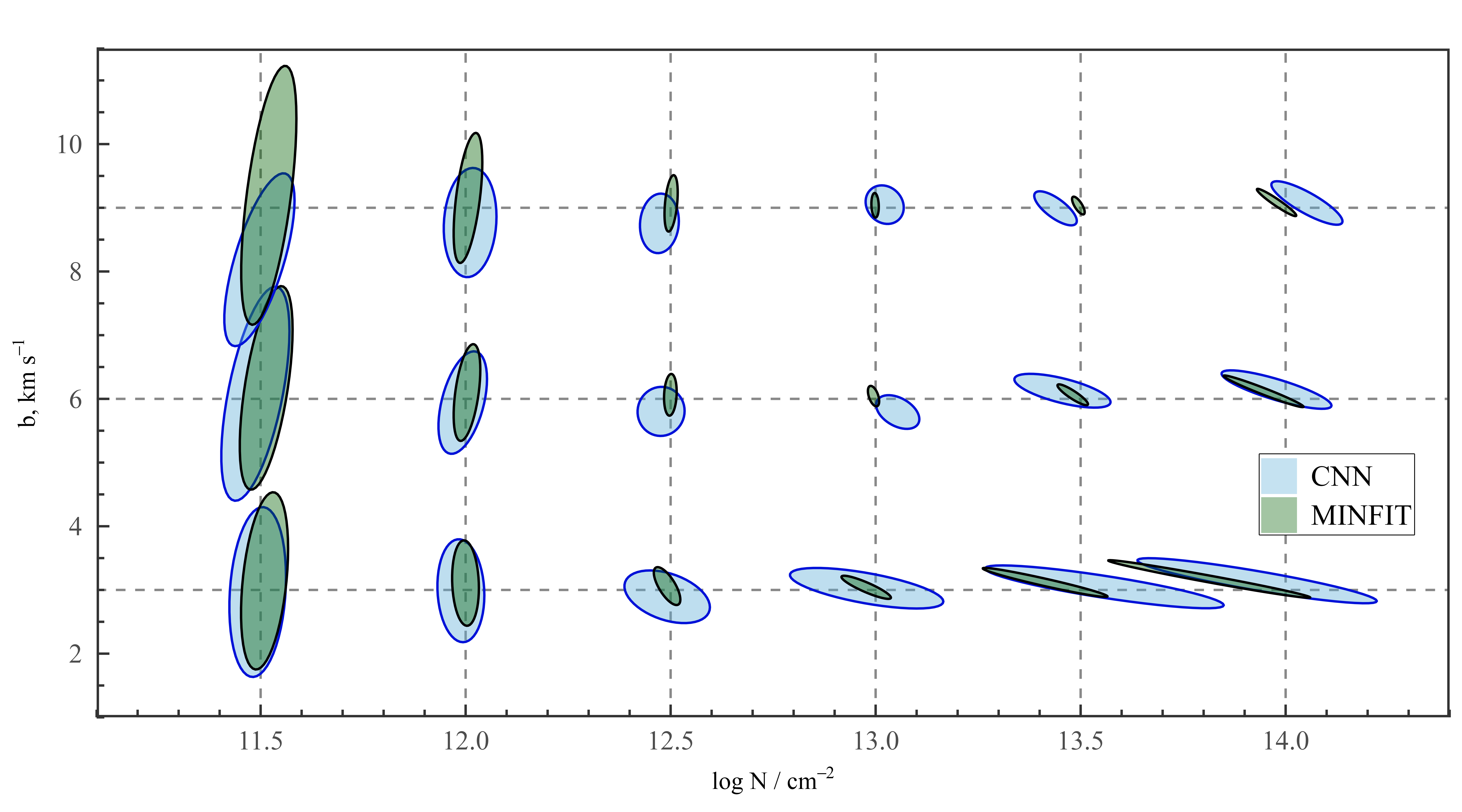}{0.95\textwidth}{}
\vspace{-15pt}
\caption{The 95\% confidence ellipses for the CNN (blue) and {\sc minfit} (green) based on $\sim~10^4$ absorption doublets with fixed $N$ and $b$ values (given by the grid intersections).  We used a flat distribution of signal-to-noise ratios in the range $\hbox{S/N} \in (9,90)$ with randomized noise realizations. The performance of both methods varies significantly across the parameter space.
}
\label{fig:variance}
\end{figure*}

We designed a CNN to predict VP parameters directly from the pixel flux values of absorption line profiles that we successfully applied to a sample of 56 single-component {\MgIIdblt} profiles observed with the HIRES and UVES spectrographs. A series of KS tests comparing the CNN predictions to those of a traditional least-squares VP fitter ({\sc minfit}) indicates that the CNN predictions are statistically indistinguishable from the VP fitter results.  Three of these tests (see Figure~\ref{fig:cdfs}) examined the distribution of the column densities, the distribution of $b$ parameters, and the distribution of $\chi^2_\nu$ values computed from the pixel-by-pixel sum of the squares of the uncertainty normalized residuals between the VP profile models and the data (see Eq.~\ref{eq:normalized-flux-residuals}).  A more stringent test of the model profile ``goodness of fit'' statistic was a comparison between the distributions of the uncertainty normalized residuals for the CNN predictions and the {\sc minfit} results (see Figure~\ref{fig:flux_residuals}).  

In addition, we note that, as gleaned from Figures~\ref{fig:method_residuals} and \ref{fig:cnn_obs}, the distribution of the residuals for the CNN predictions are comparable to those of the {\sc minfit} results and the coefficient of determination, $R^2$, reveals that less than 4\% and 7\% of the scatter in the CNN predictions of the column density and Doppler $b$ parameters, respectively, is unexplained by the model. We would point out that this unexplained scatter is on the order of the uncertainties in the {\sc minfit} parameter estimates.  Overall, the CNN predictions are statistically consistent with those derived from a traditional least-squares VP fitter, in this case {\sc minfit}.

\subsection{The Systematics of the CNN}

VP parameters are used as inputs to chemical-ionization models, from which we derive gas-phase metallicities, densities, temperatures, turbulent motions, and ionization conditions.
This means that any systematic effects resulting from VP modeling of the data can directly impact our downstream calculations.  For example, for a measured column density of ion X, we have $[{\rm X/H}] \sim \log (N_{\hbox{\tiny X}}/N_{\hbox{\tiny HI}}) -
\log (f_{\hbox{\tiny X}}/f_{\hbox{\tiny HI}})$, where {\HI} denotes neutral hydrogen, and $f_{\hbox{\tiny X}}$ and $f_{\hbox{\tiny HI}}$ are the ionization fractions of ion X and {\HI}, respectively. The ionization fractions themselves depend on the ratios of column densities, which constrain the chemical-ionization models.  Alternatively, the Doppler $b$ parameters of different ions can inform us of the gas temperatures and the degree of turbulent kinematics, as the turbulent velocity can be determined from $v_t^2 = (b^2_{\hbox{\tiny X}} - rb^2_{\hbox{\tiny Y}})/(1-r)$, where  $b_{\hbox{\tiny X}}$ and $b_{\hbox{\tiny Y}}$ are the VP component Doppler parameters for ions X and Y, and $r=m_{\hbox{\tiny Y}}/m_{\hbox{\tiny X}}$, with $r < 1$ \citep[e.g.,][]{rauch96, churchill03}.

These are but two examples of how systematic uncertainties in the measured VP column densities and $b$ parameters might skew inferences we draw from the data and, ultimately, shape our astrophysical insights.  Within the limited framework of single-component fine-structure doublet absorption, it is of interest to compare the systematic errors between the CNN predictions and the {\sc minfit} results.  In Figure~\ref{fig:variance}, we present the 95\% confidence ellipses in the VP column densities and $b$ parameters for the CNN predictions (blue) and {\sc minfit} results (green) over a grid $\log (N/{\rm cm}^{-2}) \in [11.5,14.0]$ in steps of 0.5~dex and $b=3,6,9$~{\kms}.  These ranges allow us to assess the systematic behavior across the curve of growth (see Figure~\ref{fig:cog}). Each error ellipse is measured from $10^4$ simulated {\MgII} doublets with a fixed column density and $b$ parameter employing the methods used to generate the CNN training data (HIRES and UVES spectra), including the full range of S/N (using a flat distribution).  

Inspecting Figure~\ref{fig:variance}, we see that the orientation of the confidence ellipses change with location on the curve of growth.  On the linear part, $\log (N/{\rm cm}^{-2}) \leq 12$, the uncertainty in the $b$ parameter dominates over the uncertainty in $N$.  However, for the {\sc minfit} results, the confidence ellipses are slightly more symmetric than those of the CNN.  On the other hand, for the larger $b$ (broader lines), the CNN predictions are systematically skewed toward smaller $b$.  This would indicate that for small column densities with larger $b$ values, turbulent velocities inferred from the CNN predictions would differ in a systematic way compared to the traditional VP fitting results.

On the flat part of the curve of growth, $\log (N/{\rm cm}^{-2}) \geq 13.5$, the confidence ellipses are highly elongated in the column density direction, whereas the $b$ parameters are tightly constrained.  The elongation of the confidence ellipses is most accentuated for narrower lines (smaller $b$), and they are asymmetric about the ``true'' $b$ value but with a skew toward smaller $N$ for both the CNN and {\sc minfit} at $\log (N/{\rm cm}^{-2}) = 14$.  For broader lines, there are small offsets in the column densities predicted by the CNN, and interestingly, the skews reverse direction from the narrow lines to the broader lines.  For broader lines at $\log (N/{\rm cm}^{-2}) = 13.5$, the CNN systematically slightly underpredicts $N$, but for $\log (N/{\rm cm}^{-2}) = 14$, the CNN systematically slightly overpredicts $N$. The opposite skew in $N$ occurs for the narrower lines.  The centroids of the confidence ellipses can be offset by as much as 0.1--0.3~dex.  The relative behaviors between the CNN predictions and {\sc minfit} column densities suggest that systematic difference on the order of $\sim\! 0.3$~dex in the inferred metallicities of higher column density systems would be derived between the CNN predictions and {\sc minfit} results, with the sense of the systematic offsets reflective of the confidence ellipses shown in Figure~\ref{fig:variance}.

\subsection{The Robustness of the CNN}

Supervised deep learning is highly dependent upon the quality and accuracy of the training data. In the case of spectroscopic data, by quality, we mean how well the training data capture the pixel-to-pixel noise characteristics of real-world data. By accuracy, we mean how well the training data capture the the instrumental resolution element and pixel sampling (pixels per resolution element) of real-world data.

To assess sensitivities of the trained CNN to the noise characteristics of astronomical spectroscopic data, we conducted an investigation into how the trained CNN predictions behaved as a function of the signal-to-noise ratio (S/N).  For the adopted CNN trained on the training set described in Section~\ref{subsec:training}, we tested the CNN on three copies of the validation data from Figure~\ref{fig:method_residuals}, each with a constant S/N. The samples have $\hbox{S/N}=10$, 50, and 90. We present the details of the study in Appendix~\ref{app:snr} and illustrate the residuals of the CNN predictions in Figure~\ref{fig:snr_residuals}. Summarizing, we found that the velocity and Doppler $b$ parameter predictions are robust and only mildly sensitive to the S/N of the spectra, showing a minimal increase in dispersion as S/N decreases. The column density predictions, on the other hand, were degraded for $\hbox{S/N}=10$, as up to $\sim\! 20$\% of the variance in the predictions could not be explained by the model ($R^2=0.81$). Furthermore, the tendency to underpredict the highest column density systems is more exaggerated in the $\hbox{S/N}=10$ data. The results were more robust for the $\hbox{S/N}=50$ and 90 data, returning to the performance level demonstrated in Figure~\ref{fig:method_residuals}(b) with only 2\% of the scatter was unexplained.

We also conducted a test to assess sensitivities of the trained CNN to the resolution and pixel sampling of astronomical spectroscopic data.  As described in Section~\ref{subsec:training}, the CNN was trained on $R=45,000$ spectra having $p=3$ pixels per resolution element; this yields a resolution element of $\Delta v = 6.66$~{\kms} with $\Delta v_{\rm pix} = 2.22$~{\kms}.  We then generated two copies of the $10^4$ systems shown in Figure~\ref{fig:method_residuals}, but with with different resolutions, $R=40,000$ and $R=50,000$, and tested the CNN on these samples. Our goal was to ascertain how the CNN predictions were affected when it has been trained at one resolution and is asked to make predictions on data that do not match that resolution. In other words, what if a human incorrectly trains the CNN on a slightly wrong resolution? Similarly, how robust are the CNN predictions to resolution variation across a spectrograph?  For these tests, we held the pixel sizes at $\Delta v_{\rm pix} = 2.22$~{\kms}.  For $R = 40,000$, the resolution element is $\Delta v = 6.0$~{\kms}, yielding $p=2.7$, which is a slightly smaller pixel sampling rate than the training data.  For $R = 50,000$, we have $\Delta v = 7.5$~{\kms}, yielding $p=3.5$, a slightly higher sampling rate. These tests are further detailed in Appendix \ref{app:resolution} and the results are presented in Figure~\ref{fig:resolution}.  We found that the CNN predictions were remarkably robust against discrepancies between training and test set resolution if that error is within $\sim\! 10$\% of the real-world resolution. For velocity, column density, and $b$ parameters, the percent of the scatter unexplained changed by no more than 0\%, 1\%, and 3\% respectively, and this was only for the $R=40,000$ case (when the training resolution was higher than the data resolutions).

\subsection{The Utility of the CNN}
Artificial intelligence presents an alternative approach to tackling some of our most challenging problems in astrophysics while presenting its own set of challenges. A common concern about machine learning is that the algorithms are ignorant of the underlying physics, whereas this physics is directly built into our traditional analysis algorithms. Indeed, artificial intelligence distills information, not physics.  It extracts this information directly from the data because of its superior pattern recognition, whereas traditional model-based analysis would not be designed to exploit or analyze the information unrecognized by humans.  There are many advantages of machine learning and there are many disadvantages. We refer the reader to \citet{ball10}, especially their Table~1, ``Advantages and disadvantages of well-known machine learning algorithms in astronomy.'' Regardless of the arguments on both sides of aisle, machine learning already has a rich history in the astronomical sciences and its application and methodologies are expanding as more is learned about the nuances underlying artificial intelligence \citep{smith23}.

We argue that machine learning algorithms, and in particular CNNs, are well-suited for VP decomposition of quasar absorption line systems. As we have demonstrated in this work, the VP parameters of simple doublet absorption systems are recovered with accuracies that are statistically indistinguishable from traditional VP fitting methods. However, absorption line systems, in general, are far more complex than the simple systems we analyzed.  In our application, the only physics that the CNN really navigated was the
curve of growth, which is illustrated in Figure~\ref{fig:cog}. The challenges the CNN faced were unresolved lines (small $b$ values), line saturation (high $N$ values), and S/N realizations that slightly distorted the absorption profile shapes.  This set of challenges barely covers the much broader set of challenges presented by kinematically complex absorbers with multiple transitions from multiple ions.  In addition, interloping lines from other absorbers can cause random blending, distorting the shape of one or more absorption profiles in the system.  We believe, using a step-by-step approach, that these challenges can be surmounted using supervised learning and CNNs.

An advantage of CNNs is that once trained, the CNN makes predictions at a rate of $\sim\! 20,000$ systems per minute, whereas it requires a considerable investment of time and energy for a human expert to undertake traditional VP fitting. Thus, in the case where thousands of systems require analysis, CNNs have their appeal.  Though each of the simple systems we employed in this study would require a human $\sim\! 15$ minutes to VP fit, this is not the case for more complex systems. We would note that the 422 {\MgII} absorbers VP modeled by \citet{churchill20} required 2.5 human years.  For some systems, it required 2-3 weeks to obtain a satisfactory solution.  For a duty cycle of 20 hours per week, on average, 2.5 human years equates to $\sim\! 2,500$ hours of labor.  Furthermore, simulations of multi-component VP generated synthetic {\MgII} systems indicate that traditional VP fitting fails to recover 30\% of the ``true'' VP components \citep{churchill97}.  As a consequence, the kinematics, column densities, and $b$ parameters are skewed compared to the underlying true values.  This is likely the case with real-world data. A well-trained CNN (which is by no means trivial) would be taught the ``true'' underlying distributions and may not suffer from this systematic bias. Thus, artificial intelligence holds a potential promise to more accurately inform us of the astrophysics of quasar absorption line systems. 

The disadvantage of CNNs is that very careful training is required, and training is a human intensive activity.  All permutations (kinematics, blends, multiple ions, etc.)~must be anticipated and taught to the CNN if it is to be able to generalize and make accurate predictions when faced with ``exceptions.''  Ensemble deep learning will likely be required. Thus, the bulk of the time commitment for machine learning absorption line systems lies in the design, testing, and training of CNNs. 

Traditional VP fitting methods have only grown more robust as time progresses. However, their human time-intensive commitment  threatens to render these valuable tools obsolete as the next generation of telescopes promises to increase the size of our data archives by orders of magnitude. 
Some might argue against the ``black box'' nature of machine learning algorithms. However, it is naive to pretend that traditional VP fitting is not plagued by human subjectivity. The current issue is that we are less familiar with the nuances of neural networks. But that is temporary human condition.
It is for these reasons that we must perform exploratory work such as undertaken here, so that we might characterize the behavior of artificial intelligence methods and expand our toolkit in preparation for larger and larger sets of data.

\section{Conclusion}
\label{sec:conclude}

We designed and applied a deep-learning convolutional neural network to obtain Voigt Profile (VP) models of 56 resonant fine-structure {\MgIIdblt} doublet absorption line profiles measured in HIRES \citep{vogt94} and UVES \citep{dekker00} quasar spectra ($R=45,000$).  These systems were taken from the work of \citet{churchill20}.  Using the traditional least-squares VP fitter {\sc minfit} \citep{churchill97}, they were determined to be single-component absorbers with VP parameters in the ranges 
\begin{equation}
\begin{array}{rccl}
    v &\in& (-1.5, 1.1) & {\kms} \\ 
    \log N &\in& (11.2,14.2) & {\rm cm}^{-2} \\
    b &\in& (1.6,9.6) & {\kms} \, .
\end{array}
\label{eq:conclude-vprange}
\end{equation}
Single-component doublets were selected to provide a simple control data set for this pilot study. 

The CNN was trained and run on the New Mexico State University High Performance Computing cluster \citep{trecakov21} using the TENSORFLOW package v2.9.2 \citep{abadi15}. Following a hyperparameter grid search to facilitate optimization of the supervised learning, the adopted CNN had two convolutional layers, four fully connected layers, and a 15\% dropout layer. The CNN employed ReLU activation \citep{nair10}, a learning rate of $10^{-4}$, and a mean squared error loss function coupled with an RMSprop optimizer. Training was completed after five epochs and roughly five minutes.

For training, we created  $\sim\!10^5$ simulated HIRES/UVES absorption line spectra of single-component {\MgII} doublets. We used a Latin hypercube to generate a uniform sampling of four variables per absorber, the three VP parameters ($v,N,b$) and the S/N of the spectrum.  The VP parameters bracketed the ranges given in Eq.~\ref{eq:conclude-vprange} and the S/N encompassed the range $\hbox{S/N} \in (9,90)$, which brackets the real-world data.  Via supervised learning, the CNN was taught to predict VP parameters directly from the pixel flux values of absorption line profiles.  The training was validated using a withheld sample of $10^4$ training spectra. Validation and accuracy of the CNN was assessed with two methods. (1) a regression model using the coefficient of determination, $R^2$, of the CNN predicted VP parameters versus the ``true'' known VP parameters of the withheld spectra, (2) a regression model of the CNN predicted VP parameters versus the VP parameters obtained using traditional least-squares VP fitting, which served as surrogate standard values. We then applied the CNN to the sample of 56 single-component {\MgIIdblt} absorption lines profiles. 

Summarizing our main results:

\begin{enumerate}

\item When the CNN is applied to the real-world spectra, the regression model between the CNN predictions and the {\sc minfit} results yields $R^2=0.96$ for the VP column densities, indicating that only 4\% of the scatter in the CNN predictions are unexplained by the model.  For the Doppler $b$ parameter, we obtained $R^2=0.93$, indicating that only 7\% of the scatter in the CNN predictions are unexplained by the model. The $R^2$ for the VP velocities, 0.71, would appear to suggest the CNN struggled to predict the VP velocity centers, but the scatter is on the order of 10\% of a pixel width as well as being consistent with the uncertainties in the VP velocities from {\sc minfit}.

\item  We performed a series of KS tests comparing the CNN predictions to those of a traditional least-squares VP fitter ({\sc minfit}) for the real-world spectra. We compared the distribution of (i) velocities, (ii) column densities, (iii) $b$ parameters, (iv) $\chi^2_\nu$ values computed quantifying the goodness of the fit of the predicted VP models and the data, and (v) the VP model residuals for the CNN predictions and the {\sc minfit} results. All KS tests indicated that the CNN predictions are statistically indistinguishable from the VP fitter results. 

\item  We examined the CNN performance on data with fixed noise levels (Appendix~\ref{app:snr}). We examined $\hbox{S/N} \sim 10$, 50, and 90.  We found that the predicted VP velocity is not sensitive to the S/N of the spectra.  However, the predicted VP column density for low signal-to-noise ($\hbox{S/N} \sim 10$) suffered increased scatter from the ``true'' values.  Doppler $b$ parameter predictions were slightly less accurate for low S/N.

\item  We tested the robustness of the CNN predictions under the assumption that the supervised learning employed the incorrect spectral resolution and pixel sampling rate (Appendix~\ref{app:resolution}).  We tested the trained CNN on synthetic data with two resolutions, $R$, that were $\sim\! \pm 10$\% of the resolution used for training and validation.  We found that the CNN is highly robust against mismatched spectral resolution of this degree, with only an additional 1--3\% increase in the scatter unexplained by the model for the VP column densities and Doppler $b$ parameters.

\end{enumerate}

The CNN provides statistically indistinct results from the traditional VP fitter. A caveat is that there are systematic offsets in the distributions of VP parameter predictions (see Figure~\ref{fig:variance}) that vary with location on the curve of growth ($N,b$ pairs).  However, the same is true for the traditional VP fitting software, although the sense of these systematics can differ. For example, for the $N,b$ pair $\log (N/{\rm cm}^{-2})  = 12.0$ and $b=6$~{\kms}, the CNN prediction tends to slightly underpredict both $N$ and $b$, whereas the traditional VP fitter tends to slightly overpredict $N$ and $b$. However, the 95\% confidence ellipses of the two methods overlap. That is, across the curve of growth, the magnitudes of the systematic offsets of both methods are similar, but the sense of the offsets differ. These slightly different systematic offsets between the machine learning and traditional VP fitting is a manifestation of the VP fitting problem that we have yet to fully understand.

The CNN provides results at a much faster rate than does traditional VP fitting. The generation of the training data and the supervised learning and validation of the CNN require $\sim\! 40$--60 minutes.  The CNN then analyzes the absorption systems at speeds $\sim\!10^5$ times faster than a human expert employing a traditional VP fitter.  The real time commitment for the machine learning approach to VP decomposition is twofold, (1) the design and testing of the CNN (hyperparameters), and (2) a deep understanding of the data.  The former issue is a matter of exploring and refining hyperparameters. The latter is critical, as we learned during the course of this work- even the slightest misrepresentation of the real-world data will be detected by the CNN and communicated via its predictive powers.

We emphasize that this work is not intended as a demonstration of a finalized method. One use of machine learning algorithms is to solve simpler problems as part of larger pipelines, which can often comprise ensembles of artificial intelligence methods.  Having demonstrated a simple case, we aim explore more complex absorption line systems using ensemble methods channeled through pipelines.
Our next step will be to incorporate multiple transitions from an array of low ions commonly associated with {\MgII}-selected absorbers, such as the {\FeII}, {\MnII}, {\CaII}, and {\MgI}. Such CNNs could be taught to decouple the thermal and turbulent components of the $b$ parameter. As the majority of absorption line systems are kinematically complex and have multiple VP components, we will embark on CNN designs for multi-component systems. We also aim to include further randomized complexity to our training sets, such as dead pixels, gaps in wavelength coverage, and blending from spurious absorption lines unrelated to the absorption line system of interest. Finally, we plan to experiment with newly developed methods in an attempt to obtain output uncertainties or probability distribution functions as opposed to singular prediction values.


\section*{Acknowledgments} 

The authors would like to thank Dr.~Huiping Cao for lending their expertise to help us interpret the results of and ultimately improve our CNN design. This material is based upon work supported by the National Science Foundation Graduate Research Fellowship Program under Grant No. GR0006946. Any opinions, findings, and conclusions or recommendations expressed in this material are those of the author(s) and do not necessarily reflect the views of the National Science Foundation. This work utilized resources from the New Mexico State University High Performance Computing Group, which is directly supported by the National Science Foundation (OAC-2019000), the Student Technology Advisory Committee, and New Mexico State University and benefits from inclusion in various grants (DoD ARO-W911NF1810454; NSF EPSCoR OIA-1757207; Partnership for the Advancement of Cancer Research, supported in part by NCI grants U54 CA132383 (NMSU)). SH acknowledges support for Program number HST-HF2-51507 provided by NASA through a grant from the Space Telescope Science Institute, which is operated by the Association of Universities for Research in Astronomy, incorporated, under NASA contract NAS5-26555.

\bibliographystyle{aasjournal}  
\bibliography{refs}


\appendix
\section{Noise Effects}
\label{app:snr}

To characterize how the CNN predictions may be systematically affected by the noise level in the data, we created three copies of the $\sim\!10^4$ {\MgII} doublets shown in Figure~\ref{fig:method_residuals}, each with a fixed S/N ratio. We adopted $\hbox{S/N} =10$, 50, and 90.  We will call these ``low,'' ``moderate,'' and ``high,'' respectively. We then tested our trained CNN on these data.  Recall that this CNN was trained on data with a uniform distribution of S/N ranging between 9 and 90.

\begin{figure*}[htb]
\figurenum{A1} \centering
\fig{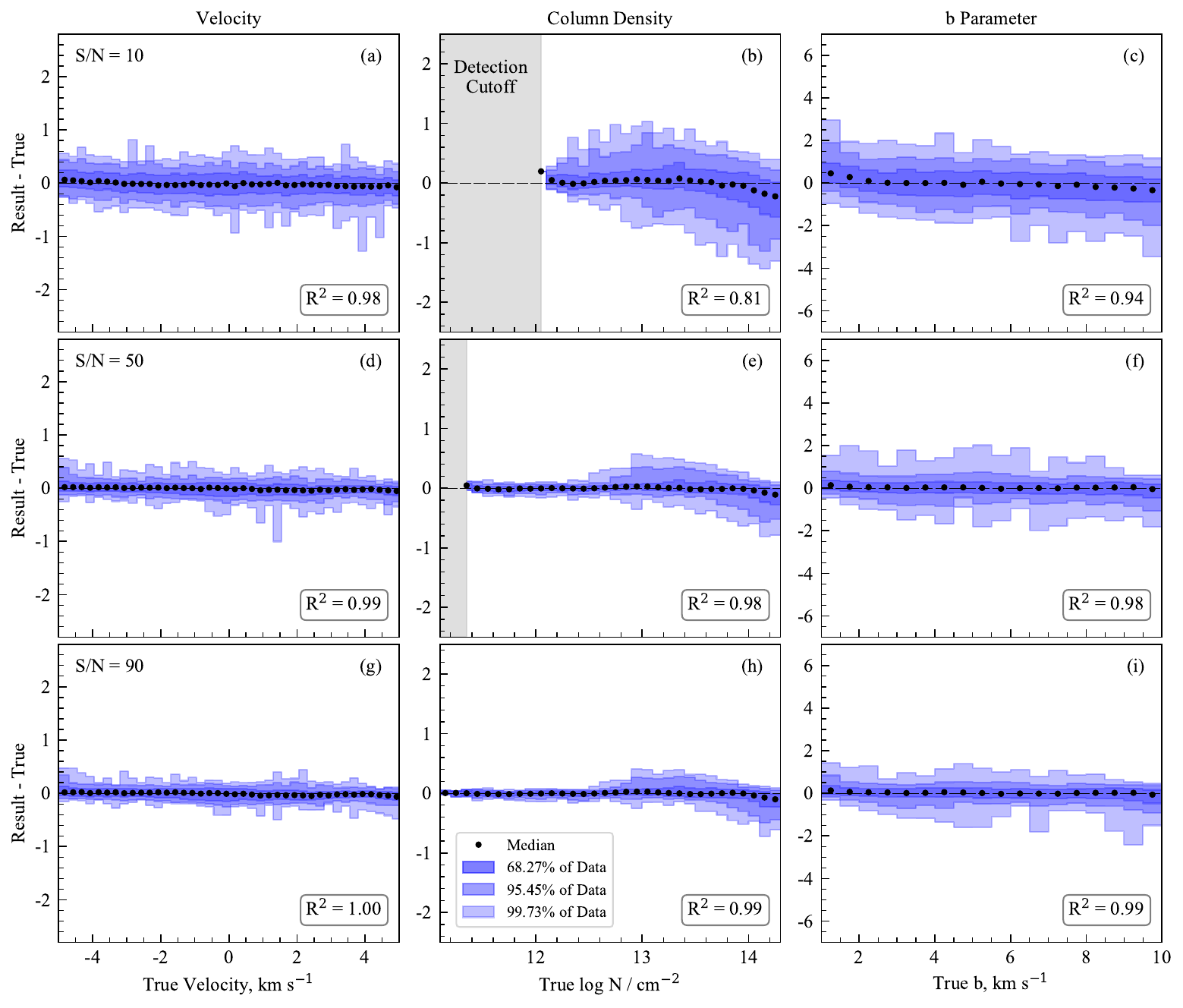}{0.85\textwidth}{}
\vspace{-15pt}
\caption{
An examination of the trained CNN performance on data sets with different fixed signal-to-noise (S/N) levels. The data presented here are described in Figure~\ref{fig:method_residuals}. (top row) low, $\hbox{S/N}=10$. (middle row) intermediate, $\hbox{S/N}=50$. (bottom row) high, $\hbox{S/N}=90$. Prediction residuals values are plotted against true values for $10^4$ spectra.  Panels from left to right show the residuals for component velocities, column densities, and $b$ parameters. No systems with column densities less than $10^{12}$ were detected in the $\hbox{S/N} = 10$ spectra or less than $10^{11.3}$ in the $\hbox{S/N} = 50$ spectra. The CNN performs well on the moderate and high S/N spectra, but shows an increase in dispersion  and slight reduction of accuracy for the low S/N spectra.
}
\label{fig:snr_residuals}
\end{figure*}

The CNN performance for the three fixed-S/N data sets is displayed as residuals between the predicted values and the true values in Figure~\ref{fig:snr_residuals}. The top row shows the prediction residuals for the low-S/N data, the middle row shows the intermediate-S/N data and the bottom row shows the high-S/N data. The left hand column is velocity, the middle is column density, and the right hand column is the Doppler $b$ parameter. Note that the smallest column density systems have not been detected in the low-S/N and intermediate-S/N spectra because the detection sensitivity drops off from $\hbox{S/N} = 50$ to $10$. As reflected in panels \ref{fig:snr_residuals}(b) and \ref{fig:snr_residuals}(e), these systems are omitted from the test as the CNN is not trained on spectra where the absorption is not detected.

If we adopt the $R^2$ coefficient as a simple measure of the accuracy of the CNN, we find that the ability of the CNN to predict the VP component velocity is fairly independent of the S/N ratio of the data.  However, the dispersion in the predictions of the Doppler $b$ parameter, and especially of the column density, increases with decreasing S/N. For the column density, $R^2>0.9$ for $\hbox{S/N} \geq 50$, but drops significantly to $R^2 \simeq 0.8$ for $\hbox{S/N} = 10$. The decline in $R^2$ for the Doppler $b$ parameter is not as significant. Examining the trends in the distribution of the prediction residuals shows there is some bias in the predictions that are enhanced in lower S/N data.  For column density, as S/N decreases, the dispersion increases and the skew in the distribution increases (toward negative residuals).  Similarly, there is a small positive skew for small Doppler $b$ parameters ($b < 2$~{\kms}) that is apparent for low-, intermediate-, and high-S/N data.  This is the regime where the absorption lines are unresolved. The positive residuals indicate that the CNN is predicting $b$ parameters that are too large by $\sim\! 1$--2~{\kms}.  A CNN faced with a system having a large column density and a small $b$ parameter would likely predict a lower column density and higher $b$ parameter, and the bias in this prediction would increase with decreasing S/N.

\section{Resolution Effects}
\label{app:resolution}

\begin{figure*}[htb]
\figurenum{B1} \centering
\fig{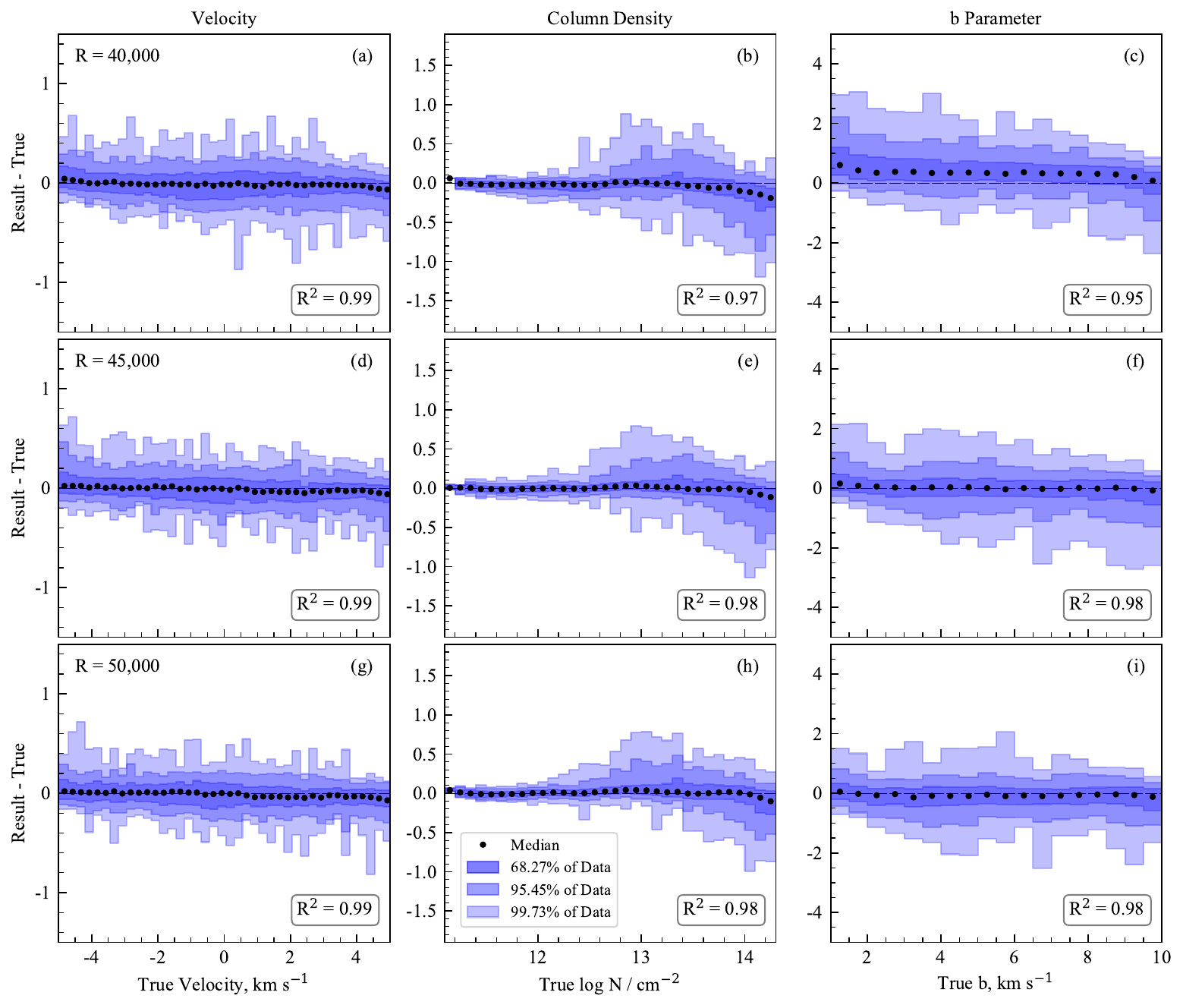}{0.85\textwidth}{}
\vspace{-15pt}
\caption{
An examination of prediction residuals between training and test data. The data presented here are described in Figure~\ref{fig:method_residuals}. The CNN was trained on R = 45,000 data and tested it on copies of the $10^4$ systems from Figure~\ref{fig:method_residuals} with different resolutions. Results minus true values are plotted against the true values. We display the results in bins of 0.25 km s$^{-1}$ for velocity, 0.1 dex for column density, and 0.5 km s$^{-1}$ for $b$ parameter. Panels from left to right show the residuals for component velocities, column densities, and $b$ parameters. Panels from top to bottom show the residuals for the R = 40,000, R = 45,000, and R = 50,000 data sets. Velocity residuals are plotted in units of pixels. The tests show that the CNN is robust against reasonable variations in resolution. The CNN shows minimal variation in results for a 10\% difference in resolution, indicating that it is robust against the smaller variations in resolution one might expect across a spectrograph.
}
\label{fig:resolution}
\end{figure*}

Even for high-resolution echelle spectrographs, spectral resolution is not perfectly constant with location on the echelle format.  This would mean that $R=\lambda/\Delta \lambda$ can vary with wavelength, or in the case of cosmologically distributed absorbers, as a function of absorber redshift.  With HIRES and UVES spectrographs, it is standard practice to assume an unchanging resolution, $R$, across the echelle format for VP fitting.  We have adopted that assumption for the training of the CNN and for the traditional VP fitting of the {\MgII} doublets studied herein.  

It is of interest to assess how sensitive CNN predictive performance is to variability in the resolution. In other words, ``What prediction biases might the CNN suffer if the CNN is trained for a fixed resolution that is different from the resolution of the data?'' To explore this question, we tested the adopted CNN, which was trained on $R=45,000$, using data with $\Delta R/R \sim \pm 0.1$ (10\% difference). We create copies of the $10^4$ absorbers from Figure~\ref{fig:method_residuals}, but with $R = 40,000$ and $R = 50,000$. This value is several factors higher than the percent variation across a free spectral range of an echelle of order $n$, i.e., $\Delta R/R = 2n/(n^2-1) \sim 0.03$ for $n=65$, a typical value of $n$. We then tested the trained CNN on these two data sets.

In Figure~\ref{fig:resolution}, we present the CNN prediction residuals. The data and layout of this figure are similar to those of Figure~\ref{fig:method_residuals} and Figure~\ref{fig:snr_residuals}, except each row displays a given resolution.  The top row is $R=40,000$, the middle row is $R=45,000$ (the same data as in Figure~\ref{fig:method_residuals}), and the bottom row is $R=50,000$.  In terms of the $R^2$ values, the clear result here is that even a 10\% difference in the adopted resolution does not diminish the ability of the CNN to make accurate predictions, nor change the characteristics of the dispersion in the distribution of prediction residuals.
We see an almost indiscernible effect for the data with higher resolution compared to the adopted resolution.  For the lower resolutions data, we do see a small decrease in the value $R^2$ compared to the adopted and higher resolution predictions.  The most salient finding is that, the for lower resolution data, the mean of the predicted Doppler $b$ parameters is skewed upward (by about 0.5--1.0~{\kms}, panel \ref{fig:resolution}(c)), and for the higher resolution data, the mean of the predicted Doppler $b$ parameters is skewed slightly downward (panel \ref{fig:resolution}(i)). This is to be expected, as lower resolution spectra will have absorption lines that are over-broadened compared to the adopted resolution and higher resolution spectra will have absorption lines that are under-broadened compared to the adopted resolution.  The CNN will clearly have its $b$ parameter predictions biased if trained on an incorrect resolution. However, we find that these biases are remarkably small for a substantial 10\% difference in resolution- an error in instrument characterization whose magnitude would be unprecedented in the literature. In summary, we find that even in the face of an exaggerated misrepresentation of the resolution of data, the CNN predictions are relatively unaltered.

\end{document}